\documentclass[apj]{emulateapj}
\usepackage{epsfig}

\def\gsim{\;\rlap{\lower 2.5pt
 \hbox{$\sim$}}\raise 1.5pt\hbox{$>$}\;}
\def\lsim{\;\rlap{\lower 2.5pt
   \hbox{$\sim$}}\raise 1.5pt\hbox{$<$}\;}
\def\ie{{\it i.e. }}
\def\eg{{\it e.g. }}
\def\gh{{\hat{g}}}
\def\Gh{{\hat{G}}}

\shorttitle{Approaching the Cram\'{e}r-Rao Bound in Weak Lensing}
\shortauthors{Zhang et al.}

\begin{document}
\title{Approaching the Cram\'{e}r-Rao Bound in Weak Lensing with PDF Symmetrization}

\author{Jun Zhang$^{1*}$, Pengjie Zhang$^1$, Wentao Luo$^2$}
\affil{$^1$Department of Physics and Astronomy, Shanghai Jiao Tong University, Shanghai 200240, China\\
$^2$Key Laboratory for Research in Galaxies and Cosmology,
Shanghai Astronomical Observatory, Shanghai 200030, China} 
\email{*betajzhang@sjtu.edu.cn}

\begin{abstract}
Weak lensing statistics is typically measured as weighted sum of shear estimators or their products (shear-shear correlation). The weighting schemes are designed in the hope of minimizing the statistical error without introducing systematic errors. It would be ideal to approach the Cram\'{e}r-Rao bound (the lower bound of the statistical uncertainty) in shear statistics, though it is generally difficult to do so in practice. The reasons may include: difficulties in galaxy shape measurement, inaccurate knowledge of the probability-distribution-function (PDF) of the shear estimator, misidentification of point sources as galaxies, etc.. Using the shear estimators defined in Zhang et al. (2015), we show that one can overcome all these problems, and allow shear measurement accuracy to approach the Cram\'{e}r-Rao bound. This can be achieved by symmetrizing the PDF of the shear estimator, or the joint PDF of shear estimator pairs (for shear-shear correlation), without any prior knowledge of the PDF. Using simulated galaxy images, we demonstrate that under general observing conditions, this idea works as expected: it minimizes the statistical uncertainty without introducing systematic error.
   
\end{abstract}


\section{Introduction}
\label{intro}

Weak gravitational lensing refers to small but coherent shape distortions (cosmic shear) of background galaxies caused by gravity. It provides a direct way of probing the cosmic structures on large scales \citep{bs01,refregier03,hj08,kilbinger15}. A number of ongoing galaxy surveys are focusing on the measurement of weak lensing statistics with a large ensemble of galaxy images, for the purpose of better understanding the cosmic evolution history and the nature of dark matter and dark energy (e.g. , DES\footnote{http://www.darkenergysurvey.org/} , HSC\footnote{http://www.naoj.org/Projects/HSC/}, KIDs\footnote{http://www.astro-wise.org/projects/KIDS/}, LSST\footnote{http://www.lsst.org/lsst}, WFIRST\footnote{http://wfirst.gsfc.nasa.gov/} ).

Currently, a great deal of efforts in the field of weak lensing are on constructing unbiased cosmic shear estimators. This is challenging due to a number of facts involved in the image formation process of modern CCD cameras, including the point spread function (PSF), the pixelation effect, the photon noise, etc.. Many different algorithms have been proposed, tested in recent open tests, and used on real galaxy data \citep{mandelbaum15}. At this stage, it is timely to raise a related question: what is the best way of taking the ensemble average of the shear estimators, or their products (for shear-shear correlation)?  

Cosmic shear is typically estimated with galaxy ellipticities. It is known that when the ensemble average of shear estimators is taken, the statistical uncertainty on the shear signal can be suppressed if larger ellipticities are weighted less, as they contribute larger shape noises than average. \citealt{bj02} (BJ02 hereafter) shows that a weighting scheme based on the probability-distribution-function (PDF) of the galaxy ellipticities can be used to achieve optimal statistical uncertainty, or the Cram\'{e}r-Rao bound (called C-R bound hereafter), given that the shear response function is properly calculated and multiplied on the ensemble average. Nevertheless, a successful application of the BJ02 idea relies on accurate measurement of galaxy ellipticities, which is unfortunately difficult in practice. For example, ellipticities estimated in the model-fitting methods generally contain biases due to noise and underfitting of galaxy morphologies \citep{vb10,bernstein10,refregier12}. 

Zhang et al.(2015) (ZLF15 hereafter) proposes an alternative form of shear estimators using the multipole moments of the galaxy power spectrum. The new method does not make assumptions on the morphologies of the galaxy or the PSF, therefore does not have the underfitting problem. The contribution of background noise to the shear estimator can be removed statistically using a neighboring noise image, and the Poisson noise contribution can also be removed directly in Fourier space. These features motivate us to understand how to approach the C-R bound with the ZLF15 shear estimators. 

Instead of taking a weighted sum, we find that one can recover the shear signal by symmetrizing the PDF of the ZLF15 shear estimators, or the joint PDF of the shear estimator pairs for measuring shear-shear correlations. It turns out that the new method allows us to approach the C-R bound without incuring systematic errors. This is realized under very general observational conditions, and without prior knowledge of the PDF. In \S\ref{MLE}, we introduce the C-R bound, and a way of realizing it through nulling of the PDF asymmetry. \S\ref{realizing_MLE_estimator} shows how to apply the PDF symmetrization method on shear estimators of ZLF15, thereby to approach the C-R bound in shear statistics, including the recovery of constant shear and shear-shear correlation. Numerical examples/proves are shown in \S\ref{numerical} using mock galaxies of very general conditions. We conclude and discuss the application of this new method in \S\ref{summary}.

\section{PDF Symmetrization Method}
\label{MLE}

\subsection{The Cram\'{e}r-Rao Bound}
\label{example}

For simplicity, let us consider $N$ random numbers $x_i$ ($i=1, 2, \cdots, N$) with an intrinsically symmetric PDF denoted as $P(x)$. Each random number is shifted by a small amount $g (\ll \sqrt{\langle x_i^2\rangle })$. Note that this situation is very similar to the case of shear estimator: $x_i$ is analogous to the galaxy ellipticity, and $g$ can be regarded as the cosmic shear signal. According to the Maximum Likelihood Estimation, an estimator $\gh$ of $g$ is given by:
\begin{equation}
\label{MLE_ave}
0=\frac{d}{d\gh}\sum_i\ln P(x_i-\gh)
\end{equation}
The C-R bound for the variance of $\gh$ is then given by:
\begin{equation}
\label{MLE_error}
\sigma_{\gh}^{-2}=-\sum_i\frac{\partial^2\ln P(x_i-\gh)}{\partial \gh^2}
\end{equation}
As a result, eq.(\ref{MLE_ave}) yields an estimate of $g$ as:
\begin{equation}
\label{MLE_ave3}
\gh=\frac{\sum_i P'(x_i)P(x_i)^{-1}}{\sum_i\left[P''(x_i)P(x_i)^{-1}-P'(x_i)^2P(x_i)^{-2}\right]}
\end{equation}
It is straightforward to show that eq.(\ref{MLE_ave3}) is unbiased. Meanwhile, the C-R bound for the variance can be derived from eq.(\ref{MLE_error}) as:
\begin{eqnarray}
\label{MLE_error2}
\sigma_{\gh}^{-2}&=&-\sum_i\frac{P''(x_i)P(x_i)-P'(x_i)^2}{P(x_i)^2}\\ \nonumber
&=&-N\int dx \frac{P''(x)P(x)-P'(x)^2}{P(x)}\\ \nonumber
&=&N\int dx \frac{P'(x)^2}{P(x)}
\end{eqnarray}
Note that here and in the rest of the paper, to simplify the notation, an integration symbol without the lower and upper limits refers to integration from negative infinity to positive infinity. The numerator in eq.(\ref{MLE_ave3}) can be regarded as a weighted sum of the data $x_i$ with the weighting function given by $P'(x_i)[P(x_i)x_i]^{-1}$, and the denominator as the sum of the weighting function multiplied by a correction factor (or response function), similar to the discussion in BJ02. However, to reach the optimal statistical uncertainty given in eq.(\ref{MLE_error2}) and an unbiased estimate of the signal $g$ simultaneously, it requires an accurate knowledge of the PDF of the data $x_i$, which is difficult if the amount of data is not large enough. More importantly, if one thinks of $x_i$ as the galaxy ellipticity plus some additional measurement errors, eq.(\ref{MLE_ave3}) becomes a biased estimator of $g$ even when the measurement error has zero mean. Due to the nonlinearity of eq.(\ref{MLE_ave3}), the correction of such a bias must be complicated. It is therefore interesting to ask if there is a way to approach the C-R bound with less stringent requirements.

\subsection{Nulling of the PDF Asymmetry} 
\label{nulling}

In the example of \S\ref{example}, $g$ causes asymmetric distribution of the measured values of $x_i$ with respect to zero. This fact suggests that one can estimate $g$ by asking how much we shall shift each $x_i$ to symmetrize the PDF of the data. To be more specific, let us set up bins for the data that are symmetrically placed around zero. We define $u_{j-1}$ and $u_j$ as the two boundaries of the $j^{th}$ bin. For convenience, we let half of the bin indices to take negative values, with $u_j=\Delta*j$ for $j=-(M-1), \cdots, -1, 0, 1, \cdots, M-1$, and $u_{\pm M}=\pm\infty$, where $\Delta$ is the bin size, and $2M$ is the total bin number. Assuming the shifted amount is $\gh$, the number of data $x_i-\gh$ that fall to the $j^{th}$ bin on the right side of $0$ is defined as:
\begin{equation}
n_j=\sum_i H(x_i-\gh-u_{j-1})H(u_j-x_i+\gh) \,\,\;(j> 0)
\end{equation}
where we have used the Heaviside step function $H$, and the subindex $i$ covers all data ID's. The bins of negative indices satisfy:
\begin{equation}
n_{-j}=\sum_i H(x_i-\gh-u_{-j})H(u_{-j+1}-x_i+\gh) \,\,\;(j> 0)
\end{equation}
Note that to avoid confusion, we explicitly write out the formulae for bins of positive and negative indices respectively.

To estimate the value of $\gh$ that can maximally symmetrize the distribution of $x_i-\gh$ with respect to zero, we form the $\chi^2$ as follows:
\begin{equation}
\label{chi2}
\chi^2=\frac{1}{2}\sum_{j> 0}\frac{(n_j-n_{-j})^2}{n_j+n_{-j}}
\end{equation}
where we assume that the fluctuation of $n_j$ obeys Poisson statistics, so that $\langle (n_j-n_{-j})^2\rangle \approx n_j+n_{-j}$. $\gh$ is estimated by minimizing $\chi^2$. Let us now show that such an estimator is unbiased. For this purpose, we assume that the number of measurements is large, so that $n_j$ and $n_{-j}$ can be written as integrations ($j> 0$):
\begin{eqnarray}
&&n_j=N_T\int_{u_{j-1}+\Delta g}^{u_j+\Delta g}dx P(x)\\ \nonumber
&&n_{-j}=N_T\int_{u_{-j}+\Delta g}^{u_{-j+1}+\Delta g}dx P(x)
\end{eqnarray} 
where $\Delta g=\gh-g$, $N_T$ is the total number of data points, and $P(x)$ is the original (symmetric) PDF of $x$ when $g=0$. With Taylor expansion, and keep terms up to the second order in $\Delta g$, we get:
\begin{eqnarray}
\frac{n_{j}}{N_T}&=&\int_{u_{j-1}}^{u_j}dx P(x)+\left[P(u_j)-P(u_{j-1})\right]\Delta g\\ \nonumber
&+&\frac{1}{2}\left[P'(u_j)-P'(u_{j-1})\right]\Delta g^2\\ \nonumber
\frac{n_{-j}}{N_T}&=&\int_{u_{-j}}^{u_{-j+1}}dx P(x)+\left[P(u_{-j+1})-P(u_{-j})\right]\Delta g\\ \nonumber
&+&\frac{1}{2}\left[P'(u_{-j+1})-P'(u_{-j})\right]\Delta g^2
\end{eqnarray} 
Since $P(x)$ is a symmetric function, we must have $P(u_j)=P(u_{-j})$ and $P'(u_j)=-P'(u_{-j})$. Therefore, 
\begin{equation}
n_j-n_{-j}=2N_T\left[P(u_j)-P(u_{j-1})\right]\Delta g
\end{equation}
Consequently, we have:
\begin{equation}
\chi^2=2N_T^2\sum_{j>0}\frac{\left[P(u_j)-P(u_{j-1})\right]^2(\gh-g)^2}{n_j+n_{-j}}
\end{equation}
which shows that when $\gh=g$, $\chi^2$ reaches its minimum, meaning that the best fit value of $\gh$ is an unbiased estimator of $g$. $\chi^2$ can be rewritten as:
\begin{equation}
\chi^2=\frac{(\gh-g)^2}{2\sigma_{\gh}^2}
\end{equation}
where 
\begin{equation}
\sigma_{\gh}^{-2}=4N_T^2\sum_{j> 0}\frac{\left[P(u_j)-P(u_{j-1})\right]^2}{n_j+n_{-j}}
\end{equation}
In the limit of small bin size $\Delta$, we have $n_j\approx n_{-j}\approx N_TP(u_j)\Delta$, therefore,
\begin{equation}
\frac{\sigma_{\gh}^{-2}}{N_T}\approx 2\sum_{j> 0}\frac{\left[P(u_j)-P(u_{j-1})\right]^2}{P(u_j)\Delta}\approx\int dx\frac{P'(x)^2}{P(x)}
\end{equation}
which recovers the C-R bound given in eq.(\ref{MLE_error2}) in the limit of small bin size. The above calculation shows that one can approach the C-R bound by symmetrizing the PDF of the data. It only requires binning the data symmetrically with respect to zero, and a reasonably large bin number. 

As examples, we consider three different types of PDF:
\begin{eqnarray}
\label{Ps}
P_1(x)&=&\frac{1}{\sqrt{2\pi}}\exp\left(-\frac{x^2}{2}\right)\\ \nonumber
P_2(x)&=&\frac{2}{\pi}(1+x^2)^{-2}\\ \nonumber
P_3(x)&=&\frac{\vert x\vert^{-2/3}}{3\sqrt{2\pi}}\exp\left(-\frac{\vert x\vert^{2/3}}{2}\right)
\end{eqnarray}
The C-R bounds (labelled as 'CR') and the variances $\sigma^2_{1,2,3}$ using the direct averaging method (labelled as 'Ave') can be worked out for the three cases respectively as:
\begin{eqnarray}
\label{sigmas}
N_T\sigma_1^2(\mathrm{Ave})&=&1,\quad N_T\sigma_1^2(\mathrm{CR})=1,\\ \nonumber
N_T\sigma_2^2(\mathrm{Ave})&=&1,\quad N_T\sigma_2^2(\mathrm{CR})=0.5,\\ \nonumber
N_T\sigma_3^2(\mathrm{Ave})&=&15,\quad N_T\sigma_3^2(\mathrm{CR})\rightarrow 0.
\end{eqnarray}

To test the PDF symmetrization method (called 'PDF-SYM' hereafter), we set the signal $g=0.01$, and use $10^7$ data points to recover the signal in each example. The results are shown in table \ref{result_test}. Note that the number in the parentheses at the end of each result refers to the statistical error on the last digit. We use this format for the notation of statistical uncertainty all through this paper. The corresponding variances $N_T\sigma^2_{1,2,3}$ in the two methods are listed in table \ref{sigma_test}. The results in the tables agree with our theoretical expectations. PDF-SYM can indeed make the statistical uncertainty approach the C-R bound when the bin number is large. In practice, $8-10$ bins are usually good enough for the purpose, unless the PDF has a number of nonmonotonic features. The figure shows that even $2$ bins can be used in PDF-SYM. This is useful when the number of data points is small ($\lsim 100$). 

Note that in making the bins, one should guarantee that each bin to have more than roughly $100$ samples, so that $\chi^2$ has a smooth dependence on the assumed value of the signal $\gh$, leading to a reliable determination of the $\chi^2$ minimum and the uncertainty of the recovered signal. The boundaries between the bins can be determined by sorting the samples according to their absolute values, and making the sample number in each bin on the positive side of zero roughly the same. The bins on the negative side are then symmetrically set up.

It is interesting to note that when the PDF has a singular behavior at the origin, such as $P_3(x)$ defined in eq.(\ref{Ps}), the C-R bound approaches zero. The results in table \ref{result_test} \& \ref{sigma_test} confirm this fact. Indeed, according to eq.(\ref{MLE_error2}), this phenomenon can occur whenever the intrinsic PDF contain sharp peaks, located either at zero, or symmetrically at the two sides of zero. This phenomenon has been previously mentioned in BJ02. It is a very useful feature in signal recovery.

As shown next, the PDF symmetrization procedure allows us to approach the C-R bound in shear measurement with shear estimators defined in ZLF15. The abovementioned advantages of the new method can be achieved under very general observational conditions.

\begin{table*}[!htb]
\centering
\caption{\textnormal{The results of signal recovery (input value is 0.01) for $10^7$ data points of three types of PDF's defined in eq.(\ref{Ps}). }}
\begin{tabular}{llllll}
\hline\hline		
Results:  &  Averaging   & PDF-SYM (2 bins)  & PDF-SYM (8 bins)     & PDF-SYM (16 bins)   & PDF-SYM (32 bins)       \\
\hline\hline
$P_1$ & 0.0102(3)  &  0.0104(4)    & 0.0101(3)    & 0.0100(4) & 0.0102(3)  \\
\hline
$P_2$ & 0.0099(3) & 0.0101(2)  & 0.0101(2)  & 0.0100(2) & 0.0101(2)\\
\hline
$P_3$ & 0.011(1)  & 0.0099999998(2) & 0.0099999998(1)  & 0.0099999998(1) & 0.0099999999(2) \\
\hline\hline		
\end{tabular}
\footnote{The number in the parentheses at the end of each result refers to the statistical error on the last digit.}
\label{result_test}
\end{table*}

\begin{table*}[!htb]
\centering
\caption{\textnormal{The measured average variances $N_T\sigma^2$ of three types of PDF's defined in eq.(\ref{Ps}).} }
\begin{tabular}{llllll}

\hline\hline		
$N_T\sigma^2$:    &  Averaging   & PDF-SYM (2 bins)  & PDF-SYM (8 bins)     & PDF-SYM (16 bins)     & PDF-SYM (32 bins)      \\
\hline\hline
$P_1$ & 1.0   &  1.6    & 1.1    & 1.2 &  0.96   \\
\hline
$P_2$ & 0.99  & 0.61  & 0.52  & 0.50 & 0.57\\
\hline
$P_3$ & 15  & $5\times 10^{-13}$ & $2\times 10^{-13}$  & $2\times 10^{-13}$ & $3\times 10^{-13}$\\
\hline\hline		
\end{tabular}
\label{sigma_test}
\end{table*}

\section{PDF-SYM in Shear Measurement}
\label{realizing_MLE_estimator}

\subsection{Shear Estimator}
\label{estimator}

Let us define the intrinsic galaxy surface brightness distribution before lensing as $f_I(\vec{x}^I)$, the lensed galaxy (before being processed by the PSF) as $f_L(\vec{x}^L)$, and the observed image as $f_O(\vec{x}^O)$, where $\vec{x}^S$ is the coordinate in the source plane, and $\vec{x}^L$ and $\vec{x}^O$ are the positions in the image plane \citep{jz11}. We have the following relations:
\begin{eqnarray}
&&f_L(\vec{x}^L)=f_I(\vec{x}^I), \quad\quad \vec{x}^I={\mathbf M}\vec{x}^L, \nonumber \\
&&f_O(\vec{x}^O)=\int d^2\vec{x}^L W_{\beta}(\vec{x}^O-\vec{x}^L)f_L(\vec{x}^L),
\label{define1}
\end{eqnarray} 
where $W_{\beta}$ is the isotropic Gaussian PSF defined as:
\begin{equation}
\label{beta}
W_{\beta}(\vec{x})=\frac{1}{2\pi\beta^2}\exp\left(-\frac{\left\vert\vec{x}\right\vert^2}{2\beta^2}\right).
\end{equation}
For now, let us only consider the case of isotropic Gaussian PSF, and no noise. ${\mathbf M}$ is the lensing distortion matrix typically defined as: ${\mathbf M}_{ij}=\delta_{ij}-\phi_{ij}$ with $\phi_{ij}=\delta_{ij}-\partial x^I_i/\partial x^L_j$ being the spatial derivatives of the lensing deflection angle. $\phi_{ij}$ is often replaced by the convergence $\kappa$ [$=(\phi_{11}+\phi_{22})/2$] and the two shear components $\gamma_1$ [$=(\phi_{11}-\phi_{22})/2$] and $\gamma_2$ ($=\phi_{12}$). The reduced shears are defined as $g_{1,2}=\gamma_{1,2}/(1-\kappa)$. The multipole moments of the galaxy power spectrum are defined as:
\begin{eqnarray}
\label{defineP}
P_{ij}&=&\int d^2\vec{k}k_1^ik_2^j\left\vert\widetilde{f_O}(\vec{k})\right\vert^2,\nonumber \\
D_n&=&\int d^2\vec{k}\left\vert\vec{k}\right\vert^n\left\vert\widetilde{f_O}(\vec{k})\right\vert^2.
\end{eqnarray}
The dependence of $P_{ij}$ on the cosmic shear can be worked out directly as: 
\begin{eqnarray}
\label{pldl}
P_{ij}&=&\vert \mathrm{det} (\mathbf{M}^{-1})\vert^2\int d^2\vec{k}k_1^ik_2^j\left\vert\widetilde{W}_{\beta}(\vec{k})\widetilde{f_I}(\mathbf{M}^{-1}\vec{k})\right\vert^2\\ \nonumber 
&=&\vert\mathrm{det} (\mathbf{M}^{-1})\vert\int d^2\vec{k}(\mathbf{M}\vec{k})_1^i(\mathbf{M}\vec{k})_2^j\left\vert\widetilde{W}_{\beta}(\mathbf{M}\vec{k})\widetilde{f_I}(\vec{k})\right\vert^2.
\end{eqnarray}
The last step is achieved by re-defining $\mathbf{M}^{-1}\vec{k}$ as $\vec{k}$. For convenience in the rest of our calculation, we define the galaxy multipole moments in the absence of lensing as:
\begin{eqnarray}
\label{pldl_I}
P_{ij}^I&=&\int d^2\vec{k}k_1^ik_2^j\left\vert\widetilde{W}_{\beta}(\vec{k})\widetilde{f_I}(\vec{k})\right\vert^2\\ \nonumber 
D_n^I&=&\int d^2\vec{k}\left\vert \vec{k}\right\vert^n\left\vert\widetilde{W}_{\beta}(\vec{k})\widetilde{f_I}(\vec{k})\right\vert^2
\end{eqnarray}
Expanding eq.(\ref{pldl}) up to the first order in shear/convergence, we get:
\begin{eqnarray}
&&P_{20}-P_{02}\\ \nonumber
&=&P_{20}^I-P_{02}^I-2g_1D_2^I+\beta^2\left[g_1D_4^I+2\kappa(P_{40}^I-P_{04}^I)\right.\\ \nonumber
&+&\left.g_1(P_{40}^I-6P_{22}^I+P_{04}^I)+4g_2(P_{31}^I-P_{13}^I)\right]
\end{eqnarray}
\begin{eqnarray}
&&2P_{11}\\ \nonumber
&=&2P_{11}^I-2g_2D_2^I+\beta^2\left[g_2D_4^I+4\kappa(P_{13}^I+P_{31}^I)\right.\\ \nonumber
&-&\left.g_2(P_{40}^I-6P_{22}^I+P_{04}^I)+4g_1(P_{31}^I-P_{13}^I)\right]
\end{eqnarray}
The shear estimators can therefore be defined as:
\begin{eqnarray}
\label{shearFourier}
\frac{1}{2}\frac{\left\langle  P_{20}-P_{02}\right\rangle }{\left\langle  D_2-\beta^2D_4/2\right\rangle }&=&-g_1,\nonumber \\
\frac{\left\langle  P_{11}\right\rangle }{\left\langle  D_2-\beta^2D_4/2\right\rangle }&=&-g_2,
\end{eqnarray}
 
The formulae are generalized in ZLF15 to take into account the conversion of the PSF form and the correction of the noise contribution. Three components ($G_1$, $G_2$, $N$) are defined as the multipole moments of the power spectrum of the galaxy image in Fourier space:
\begin{eqnarray}
\label{shear_estimator}
G_1&=&-\frac{1}{2}\int d^2\vec{k}(k_x^2-k_y^2)T(\vec{k})M(\vec{k})\\ \nonumber
G_2&=&-\int d^2\vec{k}k_xk_yT(\vec{k})M(\vec{k})\\ \nonumber
N&=&\int d^2\vec{k}\left[k^2-\frac{\beta^2}{2}k^4\right]T(\vec{k})M(\vec{k})
\end{eqnarray}
where
\begin{eqnarray}
\label{TM}
&&T(\vec{k})=\left\vert\tilde{W}_{\beta}(\vec{k})\right\vert^2/\left\vert\tilde{W}_{PSF}(\vec{k})\right\vert^2\\ \nonumber
&&M(\vec{k})=\left\vert\tilde{f}^S(\vec{k})\right\vert^2-F^S-\left\vert\tilde{f}^B(\vec{k})\right\vert^2+F^B
\end{eqnarray}
\begin{equation}
\label{shear_estimator_dis_para4}
F^S=\frac{\int_{\vert\vec{k}\vert > k_c} d^2\vec{k}\left\vert\tilde{f}^S(\vec{k})\right\vert^2}{\int_{\vert\vec{k}\vert > k_c} d^2\vec{k}}, \;\;\; F^B=\frac{\int_{\vert\vec{k}\vert > k_c} d^2\vec{k}\left\vert\tilde{f}^B(\vec{k})\right\vert^2}{\int_{\vert\vec{k}\vert > k_c} d^2\vec{k}}
\end{equation}
and $\tilde{f}^S(\vec{k})$ and $\tilde{f}^B(\vec{k})$ are the Fourier transformations of the galaxy image and a neighboring image of background noise respectively. The two additional terms $F^S$ and $F^B$ are estimates of the Poisson noise power spectra on the source and background images respectively. The critical wave number $k_c$ is chosen to be large enough to avoid the regions dominated by the source power. The factor $T(\vec{k})$ is used to convert the form of the PSF to the desired isotropic Gaussian function for correcting the PSF effect. $\beta$ should be somewhat larger than the scale radius of the original PSF to avoid singularities in the conversion. It is shown in ZLF15 that the ensemble averages of the shear estimators defined above do recover the shear values to the second order in accuracy (assuming that the intrinsic galaxy images are statistically isotropic), \ie, 
\begin{equation}
\label{shear_measure}
\frac{\left\langle  G_1\right\rangle }{\left\langle  N\right\rangle }=g_1+O(g_{1,2}^3),\;\;\;\frac{\left\langle  G_2\right\rangle }{\left\langle  N\right\rangle }=g_2+O(g_{1,2}^3)
\end{equation}
Note that the ensemble averages are taken for $G_1$, $G_2$, and $N$ separately \citep{zk11}. 

\subsection{Constant Shear Recovery with PDF-SYM}
\label{const_shear}

Eq.(\ref{shear_measure}) uses unweighted sums of the unnormalized galaxy moments. This is far from optimal, as the measurement is dominated by bright galaxies \citep{bernstein16}. Even if the shear estimators are normalized by the galaxy flux (squared), as we will show later, there is still space to further improve the statistical uncertainty. We now show that the C-R bound can be approached by symmetrizing the PDF's of $G_1$ and $G_2$.

Intuitively, according to eq.(\ref{shear_measure}), one may think that the PDF's of $G_1-g_1N$ and $G_2-g_2N$ are symmetric with respect to zero. However, as shown below, $\langle  G_i-g_iN\rangle =0$ does not guarantee that the PDF of $G_i-g_iN$ is symmetric. It is therefore necessary and interesting to dig out some details in the shear estimators of ZLF15. To do so, we should first understand the parity properties of $P_{20}-P_{02}+g_1(2D_2-\beta^2D_4)$ and $2P_{11}+g_2(2D_2-\beta^2D_4)$.

The quantities of our interests can be worked out as:
\begin{eqnarray}
\label{wrong1}
&&P_{20}-P_{02}+g_1(2D_2-\beta^2D_4)\\ \nonumber
&=&P_{20}^I-P_{02}^I+\beta^2\left[2\kappa(P_{40}^I-P_{04}^I)\right.\\ \nonumber
&+&g_1(P_{40}^I-6P_{22}^I+P_{04}^I)+\left.4g_2(P_{31}^I-P_{13}^I)\right]
\end{eqnarray}
\begin{eqnarray}
\label{wrong2}
&&2P_{11}+g_2(2D_2-\beta^2D_4)\\ \nonumber
&=&2P_{11}^I+\beta^2\left[4\kappa(P_{31}^I+P_{13}^I)\right.\\ \nonumber
&-&g_2(P_{40}^I-6P_{22}^I+P_{04}^I)+\left.4g_1(P_{31}^I-P_{13}^I)\right]
\end{eqnarray}
The above equations indicate that the PDF of $P_{20}-P_{02}+g_1(2D_2-\beta^2D_4)$ and $2P_{11}+g_2(2D_2-\beta^2D_4)$ are not exactly symmetric with respect to zero. This is due to the presence of the $P_{40}^I-6P_{22}^I+P_{04}^I$ term in both equations. The rest terms on the right sides of the equations have symmetrized PDF assuming the intrinsic galaxy images have parity symmetry statistically. It is straightforward to show that the PDF's of the following terms are symmetric:
\[P_{20}-P_{02}+g_1[2D_2-\beta^2(D_4+P_{40}-6P_{22}+P_{04})]\] and
\[2P_{11}+g_2[2D_2-\beta^2(D_4-P_{40}+6P_{22}-P_{04})]\]. In Appendix A, we show that the PDF's of these two quantities remain symmetric to the second order in shear/convergence. Note that $[P_{40}-6P_{22}+P_{04},4(P_{31}-P_{13})]$ form a pair of spin-4 quantities under spatial rotation. Their presence does not affect the ensemble average, but modifies the parity property of the PDF.

Based on the above calculation, we conclude that to use PDF-SYM, we need to define two more terms in addition to those in eq.(\ref{shear_estimator}):
\begin{eqnarray} 
\label{def_U} 
U&=&-\frac{\beta^2}{2}\int d^2\vec{k}\left(k_x^4-6k_x^2k_y^2+k_y^4\right)T(\vec{k})M(\vec{k})\\ \nonumber
V&=&-2\beta^2\int d^2\vec{k}\left(k_x^3k_y-k_xk_y^3\right)T(\vec{k})M(\vec{k})
\end{eqnarray}
It is straightforward to show that the PDF's of $G_1-g_1(N+U)$ and $G_2-g_2(N-U)$ are symmetric with respect to zero. Note that $V$ is kept for transforming $U$ in case of coordinate rotation in shear measurement.

Let us show how to measure, e.g., the first component of shear, with PDF-SYM. For convenience, let us define $B=N+U$, $G_1^S=G_1-g_1B$, and the PDF as $P_S(G_1^S,B)$, with $P_S(G_1^S,B)=P_S(-G_1^S,B)$. Note that $B$ is an observable, but $G_1^S$ is not. An observable can be defined as $\Gh_1=G_1-\gh_1B$, in which $\gh_1$ is the assumed (pseudo) value of the shear component. $\Gh_1$ is related to $G_1^S$ through:
\begin{equation}
\Gh_1=G_1^S+(g_1-\gh_1)B
\end{equation}
Define the PDF of $\Gh_1$ as $P(\Gh_1)$. We have the following relation:
\begin{eqnarray}
&&P(\Gh_1)\\ \nonumber
&=&\int dB\int dG_1^SP_S(G_1^S,B)\delta_D\left[\Gh_1-G_1^S-(g_1-\gh_1)B\right]\\ \nonumber
&=&\int dBP_S\left[\Gh_1-(g_1-\gh_1)B,B\right]
\end{eqnarray}
which is not symmetric as long as $\gh_1\ne g_1$. This fact allows us to find the unbiased estimate of $g_1$ by searching for the value of $\gh_1$ that can best symmetrize $P(\Gh_1)$. 

For this purpose, we can set up bins for $\Gh_1$'s of the galaxies that are symmetrically placed with respect to zero. The number of data that fall to the $i^{th}$ bin on the right side of zero is defined as:
\begin{equation}
n_i=\sum_j H(\Gh_1^j-u_{i-1})H(u_i-\Gh_1^j) \,\,\;(i> 0)
\end{equation}
where $u_{i-1}$ and $u_i$ are the two boundaries of the $i^{th}$ bin, and the upper index $j$ covers all galaxy ID's. Similar to the setup of \S\ref{nulling}, we let half of the bin indices to take negative values, with $u_i=-u_{-i}$, and $i=-(M-1), \cdots, -1, 0, 1, \cdots, M-1$, and $u_{\pm M}=\pm\infty$, assuming there are $2M$ bins in total. Bins of negative indices satisfy:
\begin{equation}
n_{-i}=\sum_j H(\Gh_1^j-u_{-i})H(u_{-i+1}-\Gh_1^j) \,\,\;(i> 0)
\end{equation}

To estimate the value of $\gh_1$ that can maximally symmetrize the distribution of $\Gh_1^j$ with respect to zero, we form the $\chi^2$ as follows:
\begin{equation}
\label{chi2_const}
\chi^2=\frac{1}{2}\sum_{i> 0}\frac{(n_i-n_{-i})^2}{n_i+n_{-i}}
\end{equation}
$\gh_1$ is estimated by minimizing $\chi^2$. In Appendix B, we show that {\it minimizing $\chi^2$ defined in eq.(\ref{chi2_const}) leads to an unbiased estimate of cosmic shear, with a statistical uncertainty that approaches the C-R bound in the limit of small bin sizes}.

\subsection{Shear-Shear Correlation with PDF-SYM}
\label{correlation}

Shear-shear correlation here refers to the correlation between the shear components of two galaxies defined along the line of their connection. For convenience, we use the indices $1$ and $2$ to refer to the tangential $(+)$ and cross $(\times)$ components of the shear. To use PDF-SYM, one may consider symmetrizing the PDF of the products of two shear estimators. For example, one may define the following quantity:
\begin{equation}
\xi_{11}=G_1(\vec{x})G_1(\vec{x}+\Delta\vec{x})-\hat{\xi}_{11}B(\vec{x})B(\vec{x}+\Delta\vec{x}),
\end{equation}
and use $\hat{\xi}_{11}$ that can best symmetrize the PDF of $\xi_{11}$ to infer the correlation of $g_1(\vec{x})$ and $g_1(\vec{x}+\Delta\vec{x})$ as a function of $\Delta\vec{x}$. However, it turns out that this is not a correct way, because the PDF of $g_1(\vec{x})g_1(\vec{x}+\Delta\vec{x})-\langle g_1(\vec{x})g_1(\vec{x}+\Delta\vec{x})\rangle $ is generally not symmetric with respect to zero. It turns out that we need to consider the joint PDF of the shear estimators of two galaxies.   

\begin{figure}[htbp]
\centering
\includegraphics[width=8cm,height=6cm]{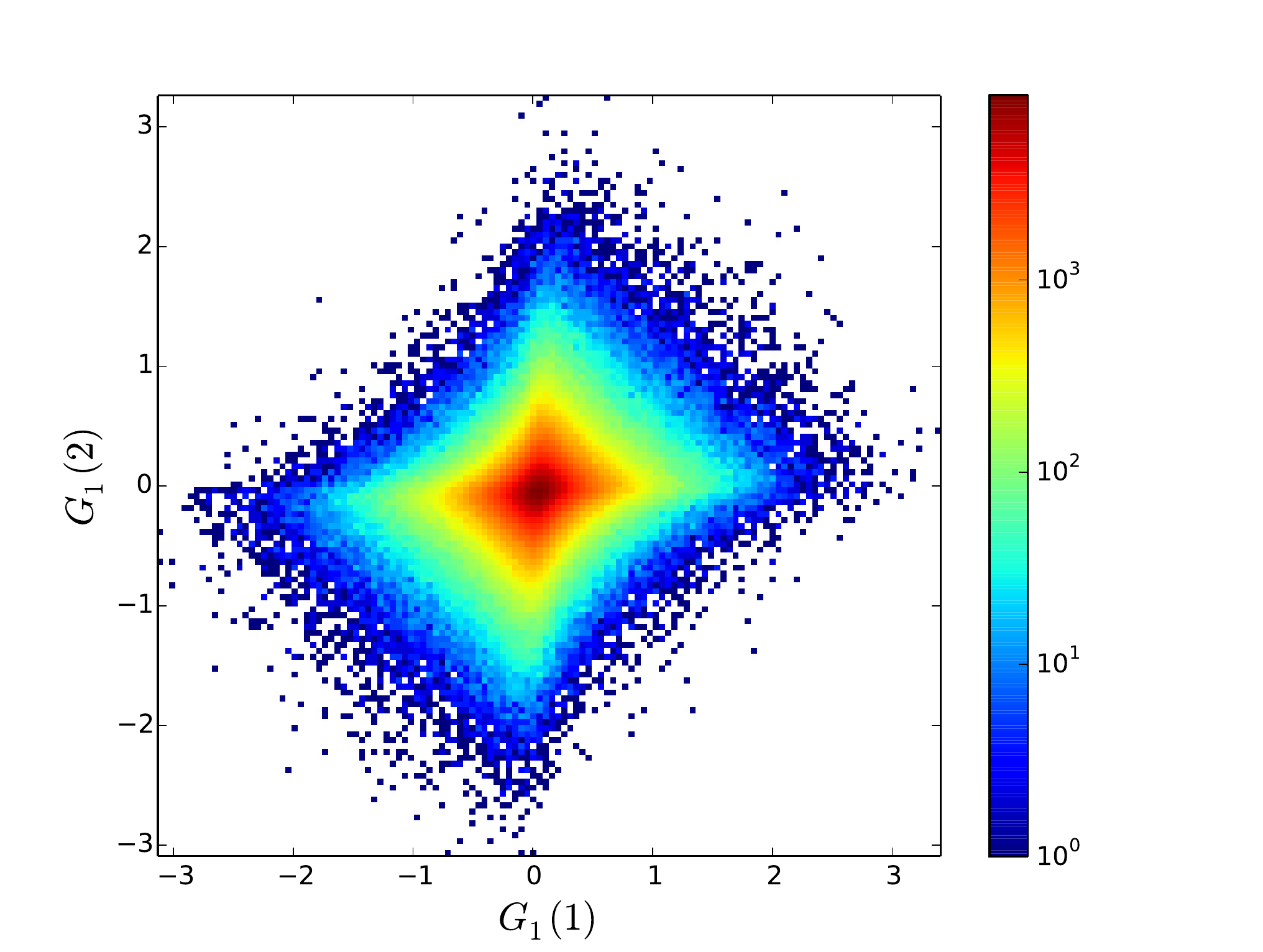}
\caption{The joint PDF of shear estimators $[G_1(1),G_1(2)]$ of two galaxies whose underlying tangential shear-components are positively correlated.}\label{joint_PDF} 
\end{figure}

In the measurement of shear-shear correlation, the shear components of galaxy pairs all have random (but correlated) values. This is different from the constant shear problem discussed in the last section. The solution is to examine the joint distribution of the correlated shear estimators, which exhibits certain asymmetric pattern. For example, fig.\ref{joint_PDF} shows the distribution of $[G_1(1),G_1(2)]$ measured from many pairs of galaxies, whose underlying shear components $[g_1(1),g_1(2)]$ satisfy a given joint Gaussian distribution with a positive correlation. The index in the parentheses refers to the galaxy ID. Note that the input shear correlation and amplitude have been amplified here for the purpose of illustration.  

We find that a way to apply PDF-SYM in the measurement of shear correlation is to apply a set of pseudo shear values generated in pairs according to an assumed Gaussian PDF of cross-correlation $\hat{\xi}$, and to find the value of $\hat{\xi}$ that can bring the joint PDF of $[G_1(1),G_1(2)]$ back to a symmetric state. We can show that the resulting value of $\hat{\xi}$ is an unbiased estimate of the opposite of the original shear-shear correlation, as it brings back the symmetry of the joint PDF. We give the details of the prove below. For convenience, we only consider the measurement of the correlation between the tangential shear components.  

Let us follow the notation of the last section. Suppose the shear estimators of two galaxies are $G_1,B,G_1',B'$, and the true underlying shear values are $g_1,g_1'$. Let us also assume that the pseudo shear values are $\gh_1,\gh_1'$, and therefore the shear estimators can be modified as: 
\begin{eqnarray}
&&\Gh_1=G_1-\gh_1B=G_1^S+(g_1-\gh_1)B \\ \nonumber
&&\Gh_1'=G_1'-\gh_1'B'={G'}_1^S+(g_1'-\gh_1')B'
\end{eqnarray}
where $G_1^S$ and ${G'}_1^S$ are the unlensed quantities that enjoy a symmetric joint PDF\footnote{Note that this point may not be true due to the presence of intrinsic alignment of galaxy shapes. In this case, one should correct the recovered shear-shear correlation by removing the intrinsic alignment contribution estimated using either close galaxy pairs or computer simulations. These topics are beyond the scope of this work.}. Define $P_S(G_1^S,B,{G'}_1^S,B')$ as the joint PDF of the unlensed quantities, which satisfy:
\begin{eqnarray}
&&P_S(G_1^S,B,{G'}_1^S,B')\\ \nonumber
&=&P_S(G_1^S,B,-{G'}_1^S,B')=P_S(-G_1^S,B,{G'}_1^S,B')
\end{eqnarray}
We can then relate the PDF of the modified shear estimators $P(\Gh_1,\Gh_1')$ to $P_S$ as:
\begin{eqnarray}
\label{PG1G1}
&&P(\Gh_1,\Gh_1')\\ \nonumber
&=&\int dg_1dg_1'\phi(g_1,g_1')\int d\gh_1d\gh_1'\hat{\phi}(\gh_1,\gh_1')\\ \nonumber
&\times&\int dB\int dB'\int dG_1^S\int d{G'}_1^SP_S(G_1^S,B,{G'}_1^S,B')\\ \nonumber
&\times&\delta_D\left[\Gh_1-G_1^S-(g_1-\gh_1)B\right]\\ \nonumber
&\times&\delta_D\left[\Gh_1'-{G'}_1^S-(g_1'-\gh_1')B'\right]\\ \nonumber
&=&\int dg_1dg_1'\phi(g_1,g_1')\int d\gh_1d\gh_1'\hat{\phi}(\gh_1,\gh_1')\int dB\int dB'\\ \nonumber
&\times&P_S\left[\Gh_1-(g_1-\gh_1)B,B,\Gh_1'-(g_1'-\gh_1')B',B'\right]
\end{eqnarray}
where $\phi(g_1,g_1')$ is the PDF of $g_1$ and $g_1'$, and $\hat{\phi}(\gh_1,\gh_1')$ is the presumed PDF of the pseudo shears $\gh_1$ and $\gh_1'$. As we will show, the form of $\hat{\phi}$ is not important (does not have to be the same as $\phi(g_1,g_1')$, but usually chosen to be Gaussian) in terms of determining the shear-shear correlation. As the shear values are small, we can Taylor expand $P_S$ to the second order in shear as:
\begin{eqnarray}
&&P_S\left[\Gh_1-(g_1-\gh_1)B,B,\Gh_1'-(g_1'-\gh_1')B',B'\right]\\ \nonumber
&=&P_S(\Gh_1,B,\Gh_1',B')-(g_1-\gh_1)B\partial_{\Gh_1}P_S\\ \nonumber
&-&(g_1'-\gh_1')B'\partial_{\Gh_1'}P_S+\frac{1}{2}(g_1-\gh_1)^2B^2\partial^2_{\Gh_1}P_S\\ \nonumber
&+&\frac{1}{2}(g_1'-\gh_1')^2B'^2\partial^2_{\Gh_1'}P_S+(g_1-\gh_1)(g_1'-\gh_1')BB'\partial_{\Gh_1}\partial_{\Gh_1'}P_S
\end{eqnarray}
Integrating over all possible values of shear, Eq.(\ref{PG1G1}) can be rewritten as:
\begin{eqnarray}
\label{PG1G1_s}
&&P(\Gh_1,\Gh_1')\\ \nonumber
&=&\int dB\int dB'\left[P_S(\Gh_1,B,\Gh_1',B')\right.\\ \nonumber
&+&\frac{1}{2}(\langle g_1^2\rangle +\langle \gh_1^2\rangle )\left(B^2\partial^2_{\Gh_1}P_S+B'^2\partial^2_{\Gh_1'}P_S\right)\\ \nonumber
&+&\left.(\langle g_1g_1'\rangle +\langle \gh_1\gh_1'\rangle )BB'\partial_{\Gh_1}\partial_{\Gh_1'}P_S\right]
\end{eqnarray}
in which we have set $\langle g_1\rangle $, $\langle g_1'\rangle $, $\langle \gh_1\rangle $, $\langle \gh_1'\rangle $, $\langle g_1\gh_1\rangle $, $\langle g_1'\gh_1'\rangle $, $\langle g_1\gh_1'\rangle $, $\langle g_1'\gh_1\rangle $ to zero, and $\langle g_1^2\rangle =\langle g_1'^2\rangle $, $\langle \gh_1^2\rangle =\langle \gh_1'^2\rangle $. On the right side of eq.(\ref{PG1G1_s}), it is clear that only the last term has odd parity, which breaks the symmetry of the joint PDF of the shear estimators. Therefore, to remove the asymmetry, we must have $\langle \gh_1\gh_1'\rangle =-\langle g_1g_1'\rangle $, which allows us to achieve an unbiased estimate of the shear-shear correlation.  
 
\begin{figure}[!htb]
\centering
\includegraphics[width=8cm,height=6cm]{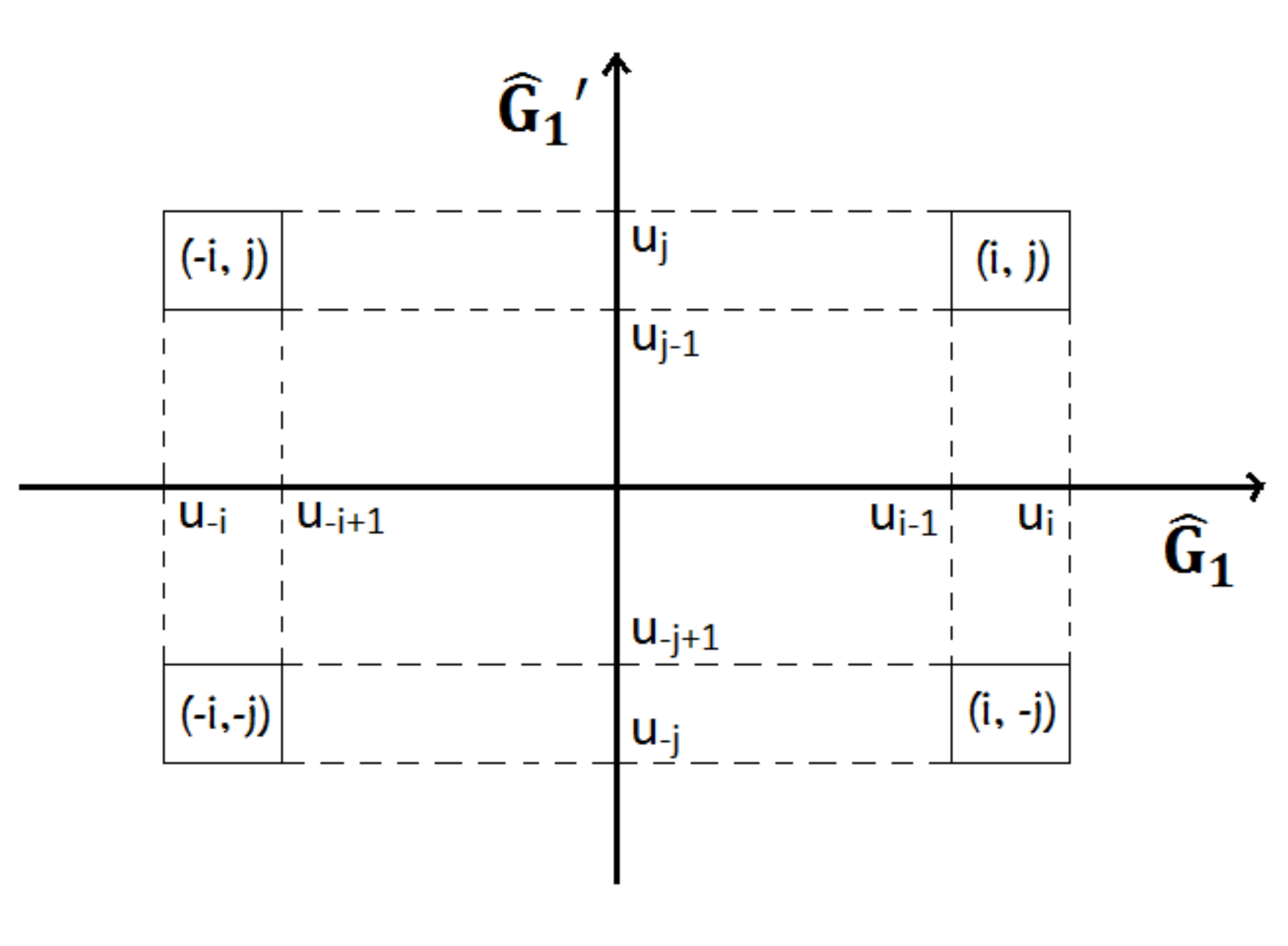}
\caption{The configuration of bins used for symmetrization of the joint PDF in shear-shear correlation measurement.}\label{bins} 
\end{figure}
For this purpose, we can set up bins that are symmetrically placed in the four quadrants in the plane of $[\Gh_1,\Gh_1']$, and each bin is labelled with two integers, as shown in fig.\ref{bins}. The number of data that fall to bin $(i,j)$ is denoted as $n_{i,j}$, and can be calculated as:
\begin{eqnarray}
n_{i,j}&=&\sum_k H(\Gh_1^k-u_{i-1})H(u_i-\Gh_1^k)\\ \nonumber
&\times&H(\Gh_1'^k-u_{j-1})H(u_j-\Gh_1'^k) \,\,\;(i,j> 0)
\end{eqnarray}
where $u_{i-1}$, $u_i$, $u_{j-1}$, $u_j$ are the boundaries of bin $(i,j)$, and the upper index $k$ is the index of galaxy pair. Again, we let half of the bin indices to take negative values, with $u_i=-u_{-i}$, and $i=-(M-1), \cdots, -1, 0, 1, \cdots, M-1$, and $u_{\pm M}=\pm\infty$, similar to the case of the last section. The number of data in bins of negative indices can be similarly defined.

To estimate the value of $\langle \gh_1\gh_1'\rangle $ that can maximally symmetrize the distribution of the joint PDF of $[\Gh_1,\Gh_1']$, we form the $\chi^2$ as follows:
\begin{equation}
\label{chi2_shear_shear}
\chi^2=\frac{1}{2}\sum_{i,j> 0}\frac{(n_{i,j}+n_{-i,-j}-n_{-i,j}-n_{i,-j})^2}{n_{i,j}+n_{-i,-j}+n_{-i,j}+n_{i,-j}}
\end{equation}
Appendix C shows that {\it minimizing $\chi^2$ leads to an unbiased estimate of $\langle g_1g_1'\rangle $ (as $-\langle \gh_1\gh_1'\rangle $), with a statistical uncertainty approaching the C-R bound in the limit of small bin sizes}.

Note that in minimizing $\chi^2$, one should fix the variances $\langle \gh_1^2\rangle$ and $\langle \gh_1'^2\rangle$ at the same value, the choice of which could be somewhat arbitrary without affecting $\chi^2$ as long as it is larger than the absolute value of the covariance $\langle \gh_1\gh_1'\rangle $. The later can be roughly estimated by the direct averaging method. Another thing to mention is that since $\gh_1$ and $\gh_1'$ are drawn randomly for a given value of $\langle \gh_1\gh_1'\rangle$, one may repeat it for several times for each $n_{i,j}$, so that the resulting $\chi^2$ is less noisy, particularly when the galaxy pair number is not large. 

Finally, we shall point out that eq.(\ref{PG1G1_s}) can be expanded to include higher order terms in shear. It is not hard to show that the next-leading-order terms that can affect the PDF symmetry of our interest is on the order of shear to the fourth power, which has been neglected here.

\section{Numerical Examples}
\label{numerical}

In this section, we present numerical examples to show the accuracy and certain characteristics of PDF-SYM. The general setup of our simulations are given in \S\ref{setup}. We discuss the case of constant shear measurement in \S\ref{constant_shear}, and shear-shear correlation in \S\ref{corr}.

\subsection{General Setup}
\label{setup}

Each of our mock galaxies is made of a number of point sources \citep{jz08}. There are a few advantages of this setup: 1) the lensing effect can be added by simply changing the positions of the point sources; 2) convolution with PSF is straightforward; 3) the richness of galaxy morphologies can be modified by changing the intrinsic distribution and the number of the point sources; 4) the image generation pipeline is very fast, suitable for testing shear recovery accuracy with a large mock galaxy ensemble. Each galaxy is placed at the center of a square grid. The pixel size of the grid is set to be the length unit in this paper. The stamp size is $48\times 48$.

The PSF has a truncated Moffat profile used in the GREAT08 project \citep{bridle09}: 
\begin{equation}
\label{PSFs}
W_{PSF}(r)\propto\left[1+\left(\frac{r}{r_d}\right)^2\right]^{-3.5}{\mathrm H}(r_c-r)
\end{equation}
The FWHM of this PSF is very close to $r_d$. We set $r_c=3r_d$ and $r_d=3$ in the simulations of this paper. In our shear measurement method, the PSF is transformed into the isotropic Gaussian form through reconvolution in Fourier space. The scale radius ($\beta$) of the target PSF is set to $r_d$, so that the size of the target PSF is somewhat larger than that of the original PSF.

\subsection{Constant Shear}
\label{constant_shear}
In this section, we study the recovery of a constant shear from a large ensemble of galaxies.

\subsubsection{Ring Galaxies}
\label{ring_gal}
In our first example, all galaxies are generated as circular rings, each of which is made of $100$ point sources (of a fixed luminosity) homogeneously placed at a fixed distance ($4$ in unit of the pixel size) from the galaxy center. The positions of the points of a galaxy are projected onto the source plane with a random projection angle, followed by the lensing and PSF effect. We generate $10000$ such galaxies, with $g_1=-0.018$ and $g_2=0.011$ ($\kappa=0$). The shear estimators are defined in eq.(\ref{shear_estimator}) and eq.(\ref{def_U}). 

\begin{table}[!htb]
\centering
\caption{\textnormal{The recovered shear values in different methods.} }
\begin{tabular}{lll}

\hline\hline		
Method                     &  $g_1 (-0.018)$                 & $g_2 (0.011)$     \\
\hline\hline
Direct Averaging & $-0.017(2)$     & $0.011(2)$     \\
\hline
PDF-SYM (2 bins) & $-0.01798(3)$     & $0.01100(2)$  \\
\hline
PDF-SYM (4 bins) & $-0.01799(2)$     & $0.01102(2)$   \\
\hline
PDF-SYM (8 bins) & $-0.01800(2)$     & $0.01101(2)$   \\
\hline\hline		
\end{tabular}
\label{result_ring}
\end{table}

\begin{figure}[!htb]
\centering
\includegraphics[width=8cm,height=6cm]{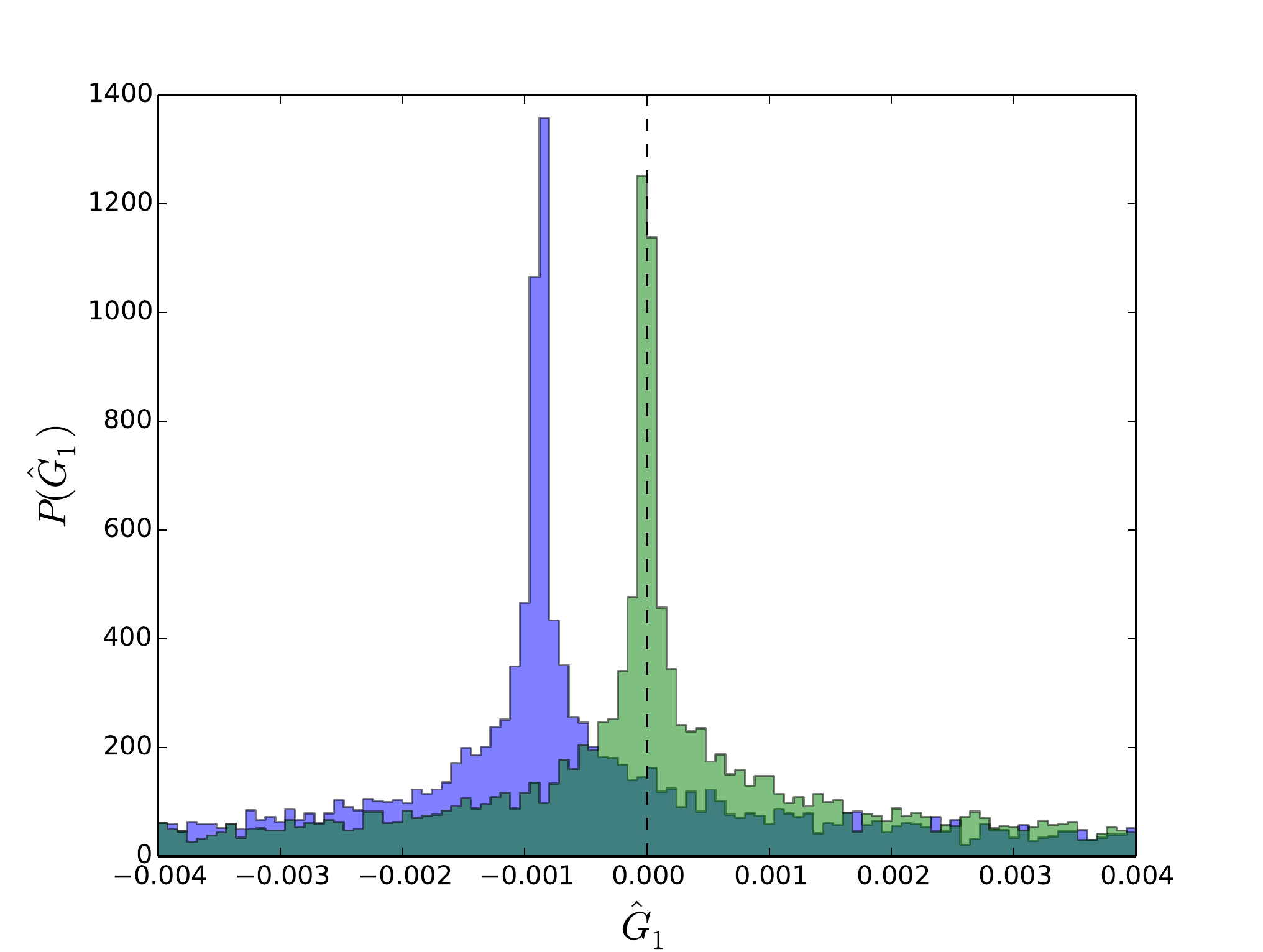}
\caption{The PDF's of $\Gh_1$ before (blue) and after (green) symmetrization for ring galaxies.}\label{perfect_ring} 
\end{figure}

Table \ref{result_ring} shows the results for four different ways of achieving the shear signals from the ensemble of shear estimators. The first method is to take the direct averages of the shear estimators, as those defined in eq.(\ref{shear_measure}). The result is shown on the first row of the table. The other three rows show the results from PDF-SYM introduced in \S\ref{const_shear}, with three different choices of bin number. In every case, the bins are symmetrically placed on the two sides of zero. The boundaries between the bins are determined by evenly dividing the galaxies according to the absolute values of $G_1$ or $G_2$, making the galaxy number in different bins roughly the same. The results in the table indicate that the PDF-SYM greatly reduces the statistical uncertainty of the recovered shear signals. Fig.\ref{perfect_ring} shows the PDF's of $\Gh_1$ before (blue) and after (green) symmetrization. It is clear that with the recovered shear value, the PDF of $\Gh_1$ is indeed symmetrized with respect to zero.  

Note that in this example, the tiny statistical uncertainty in PDF-SYM is caused by the sharp peak of the PDF at zero, as shown in \S\ref{nulling}. This is consistent with the prediction of BJ02, which discussed galaxies of pure 2D disks with random projection angles, similar to our ring galaxies. Even for the 2-bin case, the shear uncertainties in the new technique are much smaller than those from direct averaging.

\subsubsection{Mixed Types of Galaxies}
\label{mixed_gal}

In shear measurement, among the ensemble of galaxies, certain types of galaxies may be unusually sensitive to cosmic shear, such as the ring/disk galaxies with random projection angles shown in \S\ref{ring_gal}. Assuming they really exist in nature, the question is how to maximally utilize their advantages in shear recovery when their shear estimators are mixed with those of other galaxies. We now show that PDF-SYM provides a way.

Another type of galaxies we consider are generated by 2D random walks (called RW galaxies hereafter). Each galaxy is made of $100$ point sources (of constant luminosity), the positions of which are determined by a series of random walks with random directions and step sizes (between 0 and 1). When the point's position from the grid center is larger than 4, the random walk restarts from the grid center and continues from there.

\begin{figure}[!htb]
\centering
\includegraphics[width=8cm,height=6cm]{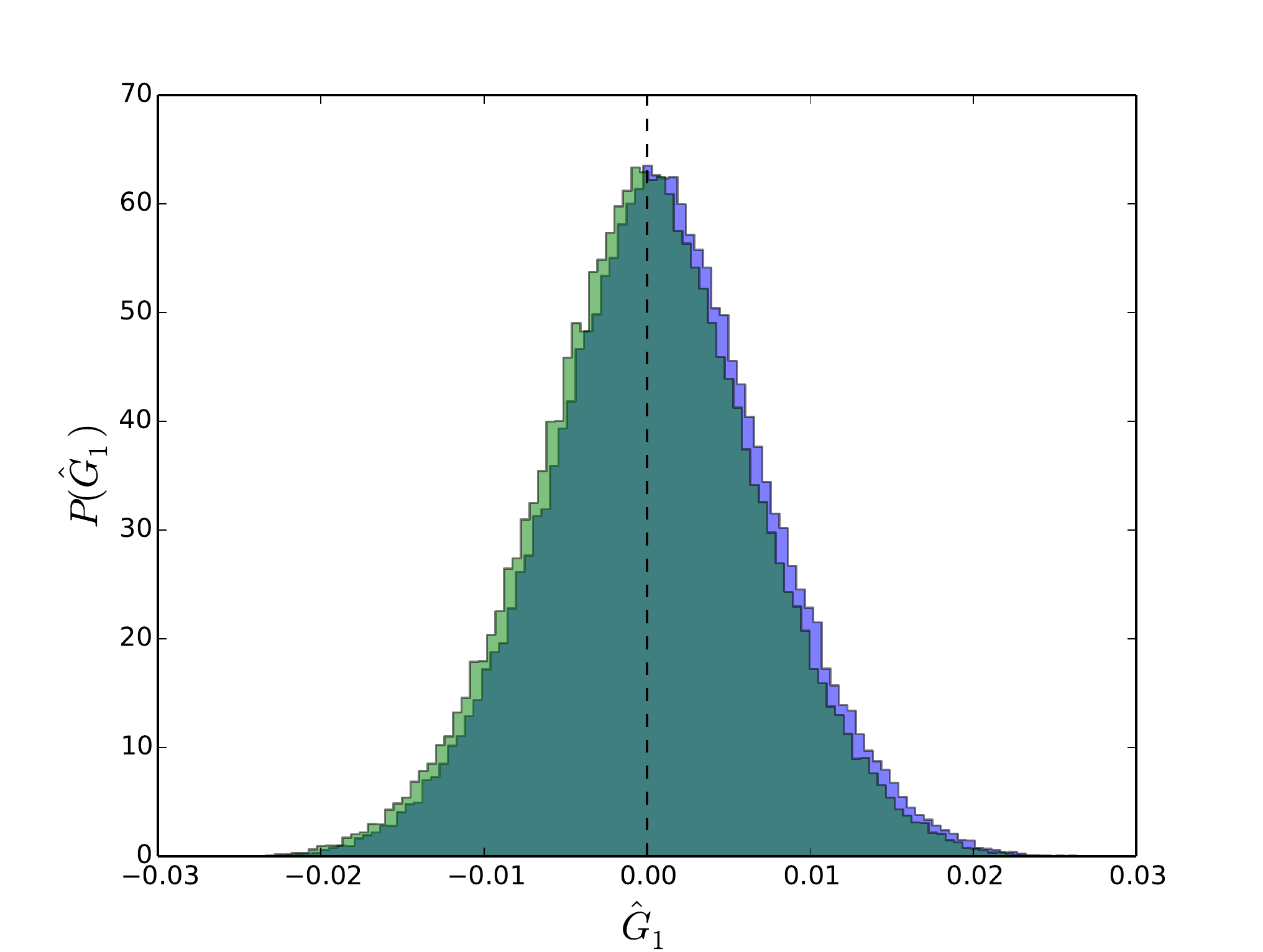}
\caption{The PDF's of $\Gh_1$ before (blue) and after (green) symmetrization for RW galaxies.}\label{pdf_rw} 
\end{figure}

The input shear values are $g_1=0.02277$, $g_2=-0.01386$ ($\kappa=0.01$). To recover shear, we generate $10^5$ RW galaxies. The results from the averaging method and PDF-SYM (8 bins) are shown in table \ref{result_mix}. For RW galaxies, PDF-SYM does not seem to improve on the statistical uncertainties with respect to the averaging method. This is very different from the case of ring galaxies. The reason is that the shear estimators' PDF shapes of the two types of galaxies are significantly different. The PDF's of $\Gh_1$ of the RW galaxies before or after symmetrization are similar to Gaussian functions, as shown in fig.\ref{pdf_rw}. They are much less peaked in the neighbourhood of zero than those of the ring galaxies shown in fig.\ref{perfect_ring}. In this case, as discussed in \S\ref{nulling}, direct averaging has a similar performance as PDF-SYM (see the example of the Gaussian PDF).   

It becomes interesting when the two types of galaxies are mixed. In the sample of RW galaxies, if $10\%$ are replaced by $10000$ ring galaxies, we find that shear recovery accuracy of the averaging method almost does not change, according to table \ref{result_mix}. In contrast, in PDF-SYM, the shear uncertainties are reduced by almost a factor of $10$, implying that the accurate shear information contained in the ring galaxies is significantly utilized. Note that this is achieved without separating the two types of galaxies.  

\begin{table*}[!htb]
\centering
\caption{\textnormal{The recovered shear values in two different methods. The input shear values are: $g_1=0.02277$, $g_2=-0.01386$.} }
\begin{tabular}{lll}
\hline\hline		
Results of $[g_1,g_2]$:  &   $10^5$ RW Galaxies          &  $9\times 10^4$ RW+$10^4$ Ring     \\
\hline\hline
 Averaging    & $[0.0226(6),-0.0130(6)]$    &   $[0.0231(6),-0.0132(6)]$     \\
\hline
PDF-SYM (8 bins) & $[0.0225(7),-0.0129(6)]$   & $[0.02278(8),-0.01392(7)]$   \\
\hline\hline 
\end{tabular}
\label{result_mix}
\end{table*}

\subsubsection{Noise and Misidentified Stars}
\label{noise_and_mis}

According to ZLF15, direct averaging of our shear estimators as defined in eq.(\ref{shear_measure}) is accurate in the presence of noise (both background noise and Poisson noise). It is also interesting to note that the accuracy is immune to possible misidentifications of stars as galaxies, because on average, point sources contribute zero values to both the numerators and the denumerators of eq.(\ref{shear_measure}). This is a very useful feature for handling faint sources. It turns out that these good characters of the shear estimators of ZLF15 remain valid in PDF-SYM. The noise and the misidentified stars make the symmetrized PDF's of the shear estimators more noisy, but do not bias the best-fit shear values. We give numerical examples below.

\begin{figure}[!htb]
\centering
\includegraphics[width=8cm,height=6cm]{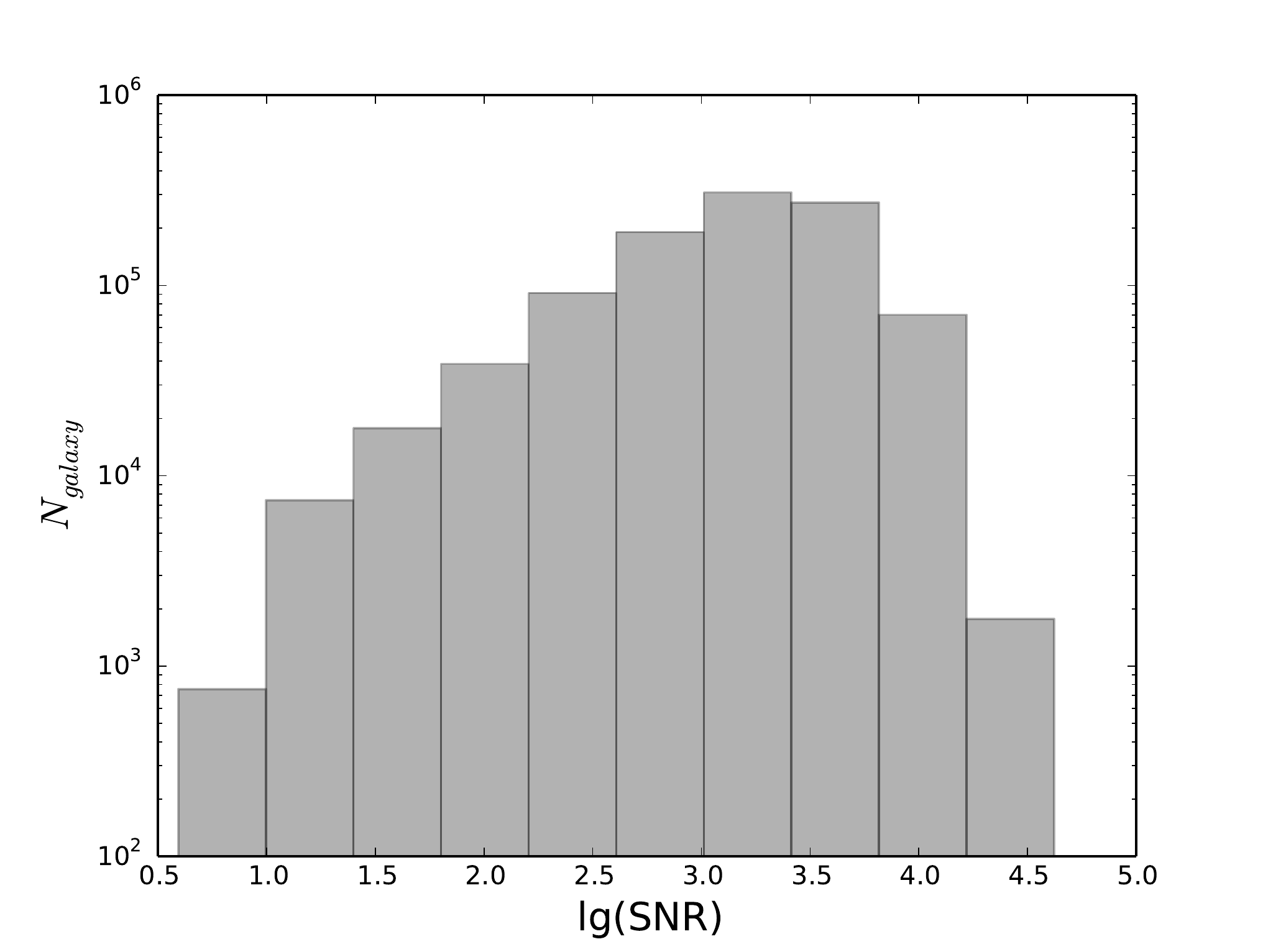}
\caption{The distribution of SNR for the galaxy images with noise used in the test of shear recovery in \S\ref{noise_and_mis}.}\label{SNR_dist} 
\end{figure}

We still use the ring and RW galaxies defined in the previous two sections. We add Poisson noise of a constant amplitude to each galaxy stamp. The total galaxy flux is random, leading to a distribution of the signal-to-noise-ratios (SNR's) of the galaxies shown in fig.\ref{SNR_dist}. $20\%$ of the sources are actually set to be stars (single-point sources). We let the stars to have the same flux distribution as the galaxies'. The input shear values are $g_1=-0.01008$, $g_2=-0.02016$ ($\kappa=-0.008$). The shear recovery results are shown in table \ref{result_noise}.

\begin{table*}[!htb]
\centering
\caption{\textnormal{Shear recovery with noise and misidentified stars. The input shear values are $g_1=-0.01008$, $g_2=-0.02016$.}}
\begin{tabular}{lllll}
\hline\hline		
Results of $[g_1,g_2]$:                    &    $10^6$ RW Galaxies            & $9\times 10^5$ RW+$10^5$ Ring     \\
\hline\hline
Averaging    & $[-0.0091(5),-0.0201(5)]$     & $[-0.0100(5),-0.0208(5)]$        \\
\hline
 PDF-SYM (8 bins) & $[-0.0101(3),-0.0207(3)]$     & $[-0.0100(1),-0.0203(1)]$   \\
\hline
 Averaging (flux normalized)   & $[-0.0102(2),-0.0203(2)]$  & $[-0.0100(2),-0.0204(2)]$ \\ 
\hline
PDF-SYM (flux normalized) & $[-0.0102(2),-0.0204(3)]$   & $[-0.01003(9),-0.02030(9)]$  \\
\hline\hline 
\end{tabular}
\label{result_noise}
\end{table*}

The results of table \ref{result_noise} indicate that through either direct averaging or PDF-SYM, shear recovery is accurate in the presence of noise and misidentified stars. The PDF method once again shows an advantage over the averaging method for the ring galaxies, and when galaxies of different types are mixed. Moreover, in this example, even for pure RW galaxies, the PDF method yields a smaller error than the averaging method. The reason is that the galaxies have a range of SNR's, and the corresponding shear estimators have very different amplitudes. Direct averaging of the shear estimators therefore under-represents the contribution from faint galaxies. One can try to weaken this problem by normalizing the shear estimators by the galaxy flux squared\footnote{Assuming the sky background has been subtracted, our galaxy flux is measured by summing over {\it the absolute values} of all the pixel readouts. This is for stablizing the total flux of the faint sources, which can be arbitrarily close to zero in principle due to the presence of noise.}. The results are shown in table \ref{result_noise} as well. Normalization does help in reducing the statistical error in the averaging method of RW galaxies, but not so much in PDF-SYM, implying that normalization of the shear estimators is not quite necessary in PDF-SYM.    

\subsubsection{GalSim Galaxies}
\label{galsim}

In this section, we further test the new method with galaxy images generated by GalSim, which is a collaborative and open-source project aiming at providing a software library for image simulations in astronomy \citep{rowe2015}. We adopt LSST like observing condition for our galaxy simulation. We set the PSF size to be 0.7", typical at the site of Cerro Tololo. For simplicity we set the PSF ellipticity fixed for the whole galaxy sample with $e_1=0.03$ and $e_2=0.02$. We simulate LSST r band disk galaxies with Sersic index equals 2, and pixel scale of 0.2". 

Two sets of simulated images are created. The first one contains $10,000$ galaxies without noise. The intrinsic galaxy ellipticities are generated so that they mimic randomly oriented disks. The shear recovery results are shown in table \ref{galsim_no_noise}. The results again confirms the advantage of the PDF-SYM method. 

\begin{figure}[!htb]
\centering
\includegraphics[width=8cm,height=6cm]{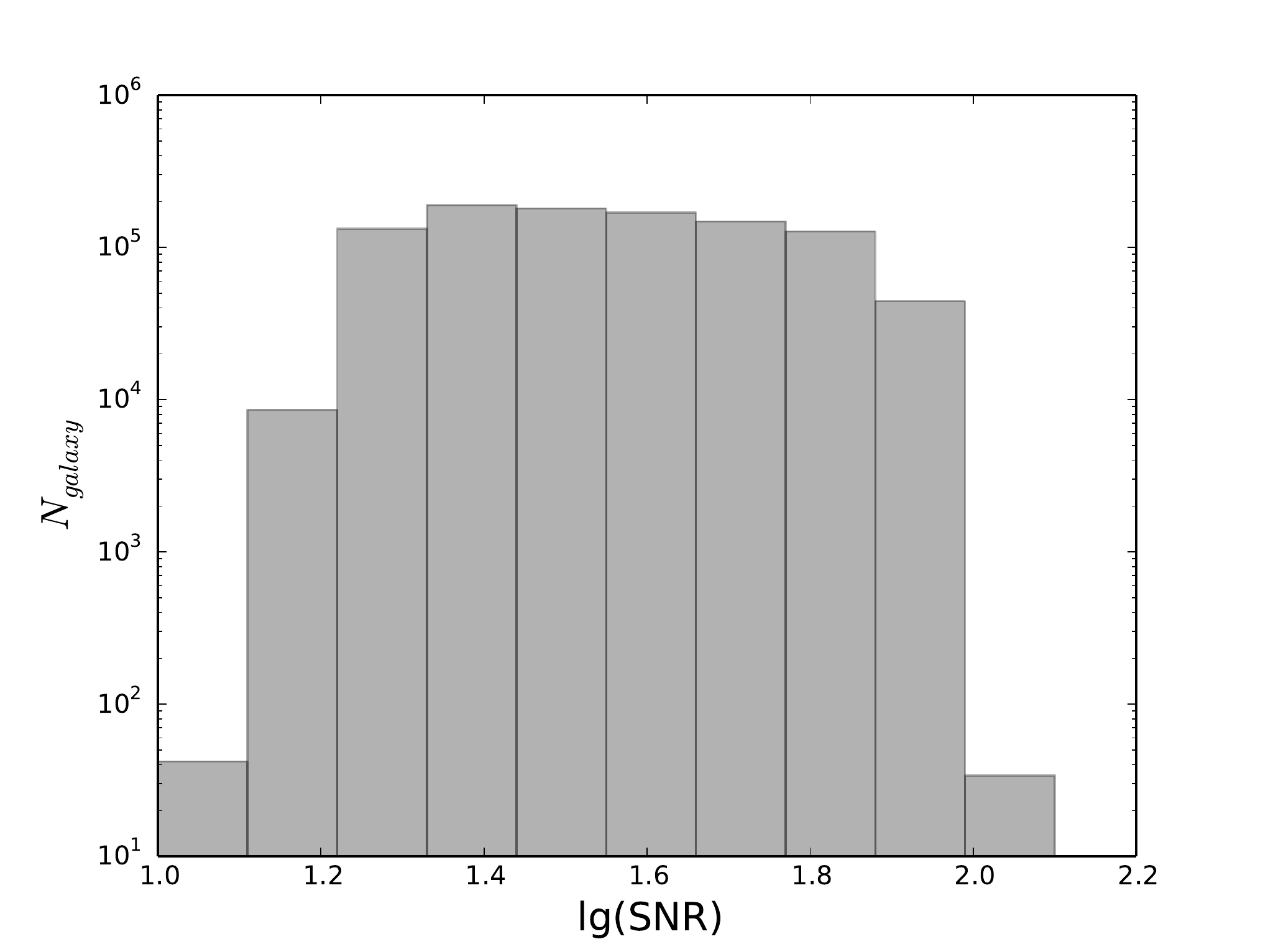}
\caption{The distribution of SNR for the galaxy images generated using GalSim in the test of shear recovery in \S\ref{galsim}.}\label{SNR_galsim} 
\end{figure}

In another example, we generate $10^6$ galaxies with SNR in the range of $20 - 100$ to check the performance of the new method for noisy galaxy images. The noise is simulated using the Exposure Time Calculator (ETC; http://lsst.org/etc). The distribution of SNR is shown in fig.\ref{SNR_galsim}. The results are shown in table \ref{galsim_with_noise}, indicating that for galaxies of smaller SNR, the performance of the PDF-SYM method becomes comparable to direct averaging. This is not surprising, as information is lost due to the existence of noise. Numerically, this is because the singular feature in the PDF of the shear estimators is smeared out by noise.

\begin{table}[!htb]
\centering
\caption{\textnormal{Shear recovery with 10000 GalSim-generated noiseless galaxies. The input shear values are $g_1=0.016$, $g_2=-0.003$.}}
\begin{tabular}{lllll}
\hline\hline		
                    &    $g_1$            & $g_2$      \\
\hline\hline
Averaging    & $0.017(1)$     & $-0.003(1)$        \\
\hline
PDF-SYM (8 bins) & $0.0162(2)$     & $-0.0029(2)$ \\ 
\hline\hline 
\end{tabular}
\label{galsim_no_noise}
\end{table}

\begin{table}[!htb]
\centering
\caption{\textnormal{Shear recovery with $10^6$ GalSim-generated galaxies with SNR in the range of $20-100$. The input shear values are $g_1=0.007$, $g_2=-0.008$.}}
\begin{tabular}{lllll}
\hline\hline		
                    &    $g_1$            & $g_2$      \\
\hline\hline
Averaging    & $0.0072(2)$     & $-0.0078(2)$        \\
\hline
PDF-SYM (8 bins) & $0.0070(2)$     & $-0.0077(2)$ \\ 
\hline\hline 
\end{tabular}
\label{galsim_with_noise}
\end{table}

\subsection{Shear-Shear Correlation}
\label{corr}

To test the recovery of shear-shear correlation, we generate a large number of galaxy pairs whose underlying shear values are correlated. To avoid cosmic variance, the shear values of each galaxy pair are not drawn from a shear field, but are generated with a coupled Gaussian distribution directly. The direction of the line connecting the two galaxies is taken to be random with respect to the grid axes. The tangential and cross components of the shear pairs are generated according to the following covariance matrix: \\

\[ \left[ \begin{array}{cc}
\langle g_t^{(1)}g_t^{(1)}\rangle  & \langle g_t^{(1)}g_t^{(2)}\rangle   \\
\langle g_t^{(2)}g_t^{(1)}\rangle  & \langle g_t^{(2)}g_t^{(2)}\rangle   \end{array} \right]
= \left[ \begin{array}{cc}
2\times 10^{-4} & 10^{-4}  \\
10^{-4} & 2\times 10^{-4} \end{array} \right],\] \\

\[ \left[ \begin{array}{cc}
\langle g_{\times}^{(1)}g_{\times}^{(1)}\rangle  & \langle g_{\times}^{(1)}g_{\times}^{(2)}\rangle   \\
\langle g_{\times}^{(2)}g_{\times}^{(1)}\rangle  & \langle g_{\times}^{(2)}g_{\times}^{(2)}\rangle   \end{array} \right]
= \left[ \begin{array}{cc}
2\times 10^{-4} & -10^{-4}  \\
-10^{-4} & 2\times 10^{-4} \end{array} \right].\] \\

Table \ref{result_corr} shows the results for the recovery of shear-shear correlations. The first three rows show the results achieved with $4\times 10^7$ galaxy pairs, and different types of galaxies and statistical methods. In these simulations, we do not add point sources as misidentified galaxies, neither any noise to the galaxy images. The last row shows the results of another simulation, in which we use $1.6\times 10^8$ galaxy pairs to recover the shear-shear correlations, with Poisson noise in every galaxy images (with SNR distribution similar to that shown in fig.\ref{SNR_dist}), and $10\%$ point sources as misidentified galaxies. We only use ring galaxies in this experiment. The results confirm the robustness of our shear measurement under general conditions, and the advantage of PDF-SYM. 

\begin{table*}[!htb]
\centering
\caption{\textnormal{The recovered shear-shear correlations. The inputs are $\langle g_t^{(1)}g_t^{(2)}\rangle=10^{-4}$ and $\langle g_{\times}^{(1)}g_{\times}^{(2)}\rangle=-10^{-4}$.}}
\begin{tabular}{lll}
\hline\hline		
Results of $[\langle g_t^{(1)}g_t^{(2)}\rangle ,\langle g_{\times}^{(1)}g_{\times}^{(2)}\rangle ]$($10^{-4}$) :                    &  Averaging                &  PDF-SYM (8$\times$8 bins)    \\
\hline\hline
$4\times 10^7$ RW Gal. Pairs & $ [1.09(8),-1.00(8)]$     & $ [1.09(8),-1.01(9)] $     \\
\hline
$4\times 10^7$ Ring Gal. Pairs & $ [1.05(7),-1.08(7)]$     & $ [1.002(5),-1.002(5)]$  \\
\hline
$4\times 10^7$ Gal. Pairs with 90\% RW and 10\% Ring & $ [1.09(8),-1.02(8)]$     & $ [0.99(3),-1.00(3)]$   \\
\hline
$1.6\times 10^8$ Ring Gal. Pairs with noise and 10\% stars  & $ [0.97(4),-1.05(4)]$     & $ [1.000(3),-1.001(3)] $     \\
\hline\hline		
\end{tabular}
\label{result_corr}
\end{table*}

\section{Summary and Discussions}
\label{summary}

Weak lensing statistics, such as mean shear or shear-shear correlation, are typically evaluated as the weighted sum of the shear estimators or their products. Traditionally, galaxy ellipticities are used as shear estimators. As discussed in BJ02, the weighting factor can be designed as a function of the galaxy ellipticities, so that the resulting statistical error can approach the Cram\'{e}r-Rao bound. This weighting scheme is however hard to realize in practice given the presence of noise and bias in the measurement of galaxy ellipticities and their PDF. 

Based on the shear estimators of ZLF15, we propose to evaluate shear statistics by symmetrizing the PDF of the shear estimator or the joint PDF of shear estimator pairs (for shear-shear correlation). We find that this is a way to approach the C-R bound without introducing systematic errors, as shown analytically in \S\ref{realizing_MLE_estimator}. 

In \S\ref{numerical}, we test shear recovery accuracy with large ensembles of galaxies or galaxy pairs (for the measurement of shear-shear correlation). We find that {\it both direct averaging and PDF symmetrization can recover shear or shear-shear correlation accurately in the presence of noise and stars that are misidentified as galaxies}, proving the robustness of both methods in practice. Note that the allowance of stars in the galaxy ensemble is a quite unusual and useful feature of the ZLF15 shear estimators. It is mostly due to the linearity of the shear estimator form.

In our numerical experiment, we use two different types of mock galaxies: 1. RW galaxies that are made of point sources connected by 2D random walks; 2. Ring galaxies that are made of point sources evenly distributed on a circle, and projected to the plane of the sky with random angles. With the method of PDF-SYM, we find that on average, every ring galaxy contains much more shear information than each RW galaxy. This is consistent with the theoretical expectation regarding the C-R bound, and agrees with BJ02, who concerns thin-disk galaxies that are similar to our ring galaxies. The advantage of ring galaxies in shear recovery cannot be easily exploited in direct averaging of the shear estimators. When these two types of galaxies are mixed in an ensemble, we find that PDF-SYM can maximally utilize the shape information in all galaxies, typically generating a much smaller statistical uncertainty than that by direct averaging. This fact is again consistent with the theory regarding the C-R bound. 

It is interesting to note that PDF-SYM allows us to approach the C-R bound in shear measurements without prior knowledge of the PDF form of the shear estimator. This is because of the monotonic dependence of the shear estimator on its corresponding shear component in the weak shear/convergence limit, implying that restoring the PDF is equivalent to symmetrizing the PDF. This fact has been proven useful for recovering the 1-point and 2-point shear statistics. Similar ideas may be developed for the measurement of n-point shear statistics in a future work. 

It is straightforward to apply the new method in several areas of weak lensing to optimize the statistical uncertainty of the results, including: 1. shear-shear correlation measurement binned in both angular separation and redshift; 2. galaxy-shear cross correlation at a fixed 3D relative position; 3. testing shear recovery accuracy. The results of these measurements all correspond to explicit theoretical expectation values, therefore are easy to interpret. In measuring shear-shear correlation (\eg, for a survey of size larger than or comparable to CFHTlens \citep{erben2013}), we find that there are typically a large number of galaxy pairs for a given angular separation and two redshift bins, making the PDF-SYM method an ideal tool to use. Note that when the number of galaxy pairs is too large, it is better to take the average of the shear estimators in each spatial bin first, and then measure the shear correlation with bin pairs of given spatial separations. We will report the application of the new method on the CFHTlens galaxy images in separate papers\footnote{If the purpose is only to make shear-map, either 2D or 3D, PDF-SYM may not be as good as direct averaging (with flux normalized shear estimators of ZLF15), because in each grid, the galaxy number may be too few ($\sim 10$ galaxies /$\mathrm{arcmin}^2$) for PDF-SYM to use more than 2 bins in shear recovery within the grid (\eg, table \ref{result_ring})}. 

If the source galaxies in the ensemble cover a broad redshift range, \eg, in the recovery of 2D shear map, or in the measurement of 2D shear-shear correlation, using PDF-SYM becomes more complicated. Indeed, in this case, the PDF of the shear estimators cannot be symmetrized at all in principle, though we can still use the $\chi^2$ formalism developed in \S\ref{const_shear} \& \S\ref{correlation}, which would result in a weighted sum of the shear signal or the shear-shear correlation within the redshift range. The weighting function can be calculated, though it requires the PDF of the shear estimator as a function of redshift. Appendix D shows a derivation of the weighting function for shear recovery. This problem will be studied more carefully in a future work.

Overall, the performance of PDF-SYM depends on the PDF form/galaxy type, distribution of galaxy SNR, and other image distortion effects \citep{rhodes10,gruen2015}. We have only examined a few cases. More work will be done to quantify the improvement with real observational data.

\acknowledgments{The authors thank Eiichiro Komatsu for motivating this work, and Zuhui Fan, Liping Fu, Guoliang Li, Chenggang Shu for useful discussions. This work is supported by the national science foundation of China (Grant No. 11273018, 11433001, 11320101002), the national basic research program of China (973 Program 2015CB857001, 2013CB834900), the national “Thousand Talents Program” for distinguished young scholars, a grant (No.11DZ2260700) from the Office of Science and Technology in Shanghai Municipal Government.}

\appendix

\section{A. Shear Estimators Accurate to the Second Order}
\label{appendix_A}

To the second order in shear/convergence, we have:

\begin{eqnarray}
&&P_{20}-P_{02}\\ \nonumber
&=&(1+2g_1^2)(P_{20}^I-P_{02}^I)+4g_1g_2P_{11}^I-2g_1D_2^I+\beta^2g_1(1-6\kappa)D_4^I+\beta^2(2\kappa-\kappa^2-5g_1^2-g_2^2)(P_{40}^I-P_{04}^I)\\ \nonumber
&+&\beta^2(1-2\kappa)\left[g_1(P_{40}^I-6P_{22}^I+P_{04}^I)+4g_2(P_{31}^I-P_{13}^I)\right]-8\beta^2g_1g_2(P_{31}^I+P_{13}^I)+2\beta^4\kappa g_1D_6^I\\ \nonumber
&+&2\beta^4g_1^2(P_{60}^I-3P_{42}^I+3P_{24}^I-P_{06}^I)+8\beta^4g_2^2(P_{42}^I-P_{24}^I)+2\beta^4\kappa^2(P_{60}^I+P_{42}^I-P_{24}^I-P_{06}^I)\\ \nonumber
&+&8\beta^4g_1g_2(P_{51}^I-2P_{33}^I+P_{15}^I)+2\beta^4\kappa \left[g_1(P_{60}^I-5P_{42}^I-5P_{24}^I+P_{06}^I)+4g_2(P_{51}^I-P_{15}^I)\right]
\end{eqnarray}

\begin{eqnarray}
&&2P_{11}\\ \nonumber
&=&(1+2g_2^2)2P_{11}^I+2g_1g_2(P_{20}^I-P_{02}^I)-2g_2D_2^I+\beta^2g_2(1-6\kappa)D_4^I+2\beta^2(2\kappa-\kappa^2-g_1^2-5g_2^2)(P_{13}^I+P_{31}^I)\\ \nonumber
&-&\beta^2(1-2\kappa)\left[g_2(P_{40}^I-6P_{22}^I+P_{04}^I)-4g_1(P_{31}^I-P_{13}^I)\right]-4\beta^2g_1g_2(P_{40}^I-P_{04}^I)+2\beta^4\kappa g_2D_6^I\\ \nonumber
&+&4\beta^4g_1^2(P_{51}^I-2P_{33}^I+P_{15}^I)+16\beta^4g_2^2P_{33}^I+4\beta^4\kappa^2(P_{51}^I+2P_{33}^I+P_{15}^I)+16\beta^4g_1g_2(P_{42}^I-P_{24}^I)\\ \nonumber
&-&2\beta^4\kappa \left[g_2(P_{60}^I-5P_{42}^I-5P_{24}^I+P_{06}^I)-4g_1(P_{51}^I-P_{15}^I)\right]
\end{eqnarray}

To the first order in shear/convergence, we have:

\begin{eqnarray}
&&2D_2-\beta^2D_4\\ \nonumber
&=&2D_2^I-\beta^2(1-6\kappa)D_4^I-2\beta^4\kappa D_6^I-4g_1(P_{20}^I-P_{02}^I)-8g_2P_{11}^I+4\beta^2\left[2g_1(P_{40}^I-P_{04}^I)+4g_2(P_{31}^I+P_{13}^I)\right]\\ \nonumber
&-&2\beta^4\left[g_1(P_{60}^I+P_{42}^I-P_{24}^I-P_{06}^I)+2g_2(P_{51}^I+2P_{33}^I+P_{15}^I)\right]
\end{eqnarray}
and
\begin{eqnarray}
&&P_{40}-6P_{22}+P_{04}\\ \nonumber
&=&(1-2\kappa)(P_{40}^I-6P_{22}^I+P_{04}^I)+2\kappa\beta^2(P_{60}^I-5P_{42}^I-5P_{24}^I+P_{06}^I)+2g_1\beta^2(P_{60}^I-7P_{42}^I+7P_{24}^I-P_{06}^I)\\ \nonumber
&-&4g_1(P_{40}^I-P_{04}^I)+4g_2\beta^2(P_{51}^I-6P_{33}^I+P_{15}^I)+8g_2(P_{31}^I+P_{13}^I)
\end{eqnarray}

Therefore, we have:
\begin{eqnarray}
&&P_{20}-P_{02}+g_1\left[2D_2-\beta^2\left(D_4+P_{40}-6P_{22}+P_{04}\right)\right]\\ \nonumber
&=&(1-2g_1^2)(P_{20}^I-P_{02}^I)-4g_1g_2P_{11}^I+\beta^2\left[(2\kappa-\kappa^2+7g_1^2-g_2^2)(P_{40}^I-P_{04}^I)+4g_2(1-2\kappa)(P_{31}^I-P_{13}^I)\right]\\ \nonumber
&+&\beta^4\left[-2g_1^2(P_{60}^I-3P_{42}^I+3P_{24}^I-P_{06}^I)+8g_2^2(P_{42}^I-P_{24}^I)+2\kappa^2(P_{60}^I+P_{42}^I-P_{24}^I-P_{06}^I)+8\kappa g_2(P_{51}^I-P_{15}^I)\right],
\end{eqnarray}

\begin{eqnarray}
&&2P_{11}+g_2\left[2D_2-\beta^2\left(D_4-P_{40}+6P_{22}-P_{04}\right)\right]\\ \nonumber
&=&(1-2g_2^2)2P_{11}^I-2g_2g_1(P_{20}^I-P_{02}^I)+\beta^2\left[(2\kappa-\kappa^2-g_1^2+7g_2^2)2(P_{31}^I+P_{13}^I)+4g_1(1-2\kappa)(P_{31}^I-P_{13}^I)\right]\\ \nonumber
&+&\beta^4\left[-16g_2^2P_{33}^I+4g_1^2(P_{51}^I-2P_{33}^I+P_{15}^I)+4\kappa^2(P_{51}^I+2P_{33}^I+P_{15}^I)+8\kappa g_1(P_{51}^I-P_{15}^I)\right]
\end{eqnarray}

It is easy to show that the PDF's of the right sides of the above equations are symmetric with respect to zero assuming the intrinsic galaxy images have parity symmetry.

\section{B. PDF-SYM for Constant Shear}
\label{appendix_B}

Let us now show that minimizing $\chi^2$ defined in eq.(\ref{chi2_const}) yields an estimate of shear that is unbiased, and the statistical uncertainty approaches the C-R bound in the limit of small bin sizes. For this purpose, we assume that the number of measurements is large, so that $n_i$ and $n_{-i}$ can be written as integrations ($i> 0$):

\begin{equation}
n_i=N_T\int_{u_{i-1}}^{u_i}d\Gh_1P(\Gh_1)=N_T\int dB\int_{u_{i-1}}^{u_i}d\Gh_1P_S\left[\Gh_1-(g_1-\gh_1)B,B\right]=N_T\int dB\int_{u_{i-1}+B\Delta g }^{u_i+B\Delta g }dA P_S(A,B)
\end{equation} 
where $\Delta g=\gh_1-g_1$. Similarly, we have: 
\begin{equation}
n_{-i}=N_T\int dB\int_{u_{-i}+B\Delta g }^{u_{-i+1}+B\Delta g}dA P_S(A,B)
\end{equation} 
Since $P_S(A,B)$ is an even function with respect to $A$, we must have $P_S(u_i,B)=P_S(u_{-i},B)$ and $\partial_AP_S(A,B)\vert_{A=u_i}=-\partial_AP_S(A,B)\vert_{A=u_{-i}}$. Therefore, 
\begin{equation}
n_i-n_{-i}=2N_T\int dB\left[P_S(u_i,B)-P_S(u_{i-1},B)\right]B\Delta g
\end{equation}
Consequently, we have:
\begin{equation}
\chi^2=N_T(\gh_1-g_1)^2\sum_{i> 0}\frac{\left\{\int dB\left[P_S(u_i,B)-P_S(u_{i-1},B)\right]B\right\}^2}{\int dB\int_{u_{i-1}}^{u_i}dA P_S(A,B)}
\end{equation}
which shows that when $\gh_1=g_1$, $\chi^2$ reaches its minimum, meaning that the best fit value of $\gh_1$ is an unbiased estimator of $g_1$. $\chi^2$ can be rewritten as:
\begin{equation}
\chi^2=\frac{(\gh_1-g_1)^2}{2\sigma_{\gh_1}^2}
\end{equation}
In the limit of small bin size $\Delta$, we have:
\begin{equation}
\frac{\sigma_{\gh_1}^{-2}}{N_T}\approx 2\sum_{i> 0}\frac{\left\{\int dB\left[P_S(u_i,B)-P_S(u_{i-1},B)\right]B\right\}^2}{\int dB P_S(u_i,B)\Delta}\approx \int dA\frac{\left[\int dB\partial_AP_S(A,B)B\right]^2}{\int dB P_S(A,B)}
\end{equation}

In comparison, let us work out the C-R bound of shear. The PDF of $G_1$ for a $\gh_1$ is:
\begin{equation}
P(G_1)=\int dB\int dG_1^SP_S(G_1^S,B)\delta_D(G_1-g_1B-G_1^S)=\int dBP_S(G_1-g_1B,B)
\end{equation}
So that:
\begin{equation}
\sigma_{g_1}^{-2}(MLE)=-\sum_i\frac{\partial^2\ln P(G_1^i) }{\partial g_1^2}=N_T\int dA \frac{\left[\int dB B \partial_AP_S(A,B)\right]^2}{\int dB P_S(A,B)}
\end{equation}
which agrees with the result of PDF-SYM in the limit of small bin size.

\section{C. PDF-SYM for Shear-Shear Correlation}
\label{appendix_C}

Assuming the number of measurements is large, so that $n_{i,j}$ can be expressed as integrations:
\begin{equation}
n_{i,j(> 0)}=N_T\int_{u_{i-1}}^{u_i}d\Gh_1\int_{u_{j-1}}^{u_j}d\Gh_1'P(\Gh_1,\Gh_1')
\end{equation}
Using eq.(\ref{PG1G1_s}), and the parity properties of the function $P_S$, we can show:
\begin{eqnarray}
&&(n_{i,j}+n_{-i,-j}-n_{-i,j}-n_{i,-j})^2=16N_T^2(\langle g_1g_1'\rangle +\langle \gh_1\gh_1'\rangle )^2\\ \nonumber
&\times&\left\{\int dB\int dB'(BB')\left[P_S(u_i,B,u_j,B')-P_S(u_i,B,u_{j-1},B')-P_S(u_{i-1},B,u_j,B')+P_S(u_{i-1},B,u_{j-1},B')\right]\right\}^2
\end{eqnarray}
and 
\begin{equation}
n_{i,j}+n_{-i,-j}+n_{-i,j}+n_{i,-j}=4N_T\int dB\int dB'\int_{u_{i-1}}^{u_i}dA \int_{u_{j-1}}^{u_j}dA' P_S(A,B,A',B')
\end{equation}
Consequently, we have:
\begin{eqnarray}
\chi^2&=&2N_T(\langle g_1g_1'\rangle +\langle \gh_1\gh_1'\rangle )^2\sum_{i,j> 0}\left[\int dB\int dB'\int_{u_{i-1}}^{u_i}dA \int_{u_{j-1}}^{u_j}dA' P_S(A,B,A',B')\right]^{-1}\\ \nonumber
&\times&\left\{\int dB \int dB'(BB')\left[P_S(u_i,B,u_j,B')-P_S(u_i,B,u_{j-1},B')-P_S(u_{i-1},B,u_j,B')+P_S(u_{i-1},B,u_{j-1},B')\right] \right\}^2
\end{eqnarray}

When $\langle \gh_1\gh_1'\rangle =-\langle g_1g_1'\rangle $, $\chi^2$ reaches its minimum. $\chi^2$ can be rewritten as:
\begin{equation}
\chi^2=\frac{(\langle \gh_1\gh_1'\rangle +\langle g_1g_1'\rangle )^2}{2\sigma_{\langle \gh_1\gh_1'\rangle }^2}
\end{equation}
In the limit of small bin size $\Delta$, we have:
\begin{eqnarray}
\frac{\sigma_{\langle \gh_1\gh_1'\rangle }^{-2}}{N_T}&\approx&4\Delta^2\sum_{i,j> 0}\left[\int dB \int dB'(BB')\partial_{u_i}\partial_{u_j}P_S(u_i,B,u_j,B')\right]^2\left[\int dB\int dB'P_S(u_i,B,u_j,B')\right]^{-1}\\ \nonumber
&\approx&\int dA\int dA'\frac{\left[\int dB \int dB'(BB')\partial_{A}\partial_{A'}P_S(A,B,A',B')\right]^2}{\int dB\int dB'P_S(A,B,A',B')}
\end{eqnarray}

In comparison, let us work out the C-R bound for the shear-shear correlation. The PDF of $G_1$ for a $\gh_1$ is:
\begin{eqnarray}
P(G_1,G_1')
&=&\int dg_1 \int dg_1' \phi(g_1,g_1')\int dB\int dB'P_S(G_1-g_1B,B,G_1'-g_1'B',B')\\ \nonumber
&=&\int dB\int dB'\left[P_S(G_1,B,G_1',B')+\frac{1}{2}\langle g_1^2\rangle B^2\partial_{G_1}^2P_S+\frac{1}{2}\langle {g_1'}^2\rangle B'^2\partial_{G_1'}^2P_S+\langle g_1g_1'\rangle BB'\partial_{G_1}\partial_{G_1'}P_S\right]
\end{eqnarray}
So that:
\begin{equation}
\sigma_{\langle g_1g_1'\rangle }^{-2}(MLE)=-\sum_i\frac{\partial^2\ln P(G_1^i,{G_1'}^i) }{\partial {\langle g_1g_1'\rangle }^2}=N_T\int dA\int dA'\frac{\left[\int dB \int dB'(BB')\partial_{A}\partial_{A'}P_S(A,B,A',B')\right]^2}{\int dB\int dB'P_S(A,B,A',B')}
\end{equation}
which again agrees with the result of PDF-SYM in the limit of small bin size.

\section{D. PDF-SYM for Recovery of Shear covering a redshift range}
\label{appendix_D}

Let us consider the case in which the source galaxies inside a given angular region covers a certain redshift range. Minimizing $\chi^2$ defined in eq.(\ref{chi2_const}) then leads to an estimate of a weighted sum of the shear signal along the line of sight. The weighting function can be calculated straightforwardly. For this purpose, we still assume that the number of measurements is large, so that $n_i$ can be written as integrations ($i> 0$):

\begin{equation}
n_i=N_T\int_{u_{i-1}}^{u_i}d\Gh_1P(\Gh_1)=N_T\int dz\int dB\int_{u_{i-1}}^{u_i}d\Gh_1P_S\left[\Gh_1-(g_1(z)-\gh_1)B,B,z\right]
\end{equation} 
Note that we need to denote the PDF $P_S$ as a function of redshift in this case, because the image qualities of the observed galaxies must have systematic dependence on the redshift. The difference between $n_i$ and $n_{-i}$ can be written as: 
\begin{equation}
n_i-n_{-i}=2N_T\int dz\left[\gh_1-g_1(z)\right]\int dB B\left[P_S(u_i,B,z)-P_S(u_{i-1},B,z)\right]
\end{equation}
Consequently, we get:
\begin{equation}
\chi^2=N_T\sum_{i> 0}\frac{\left\{\int dz \left[\gh_1-g_1(z)\right]\int dB B\left[P_S(u_i,B,z)-P_S(u_{i-1},B,z)\right]\right\}^2}{\int dz\int dB\int_{u_{i-1}}^{u_i}dA P_S(A,B,z)}
\end{equation}
from which one can show that when $\chi^2$ reaches its minimum, \ie, when $d\chi^2/d\gh_1=0$, the resulting $\gh_1$ corresponds to the following weighted sum of $g_1(z)$:
\begin{equation}
\gh_1=\int dz g_1(z)\omega(z)
\end{equation}
where
\begin{equation}
\omega(z)=\left(\frac{1}{\sum_{i>0}E_i^2F_i^{-1}}\right)\sum_{i>0}E_iF_i^{-1}\int dB B\left[P_S(u_i,B,z)-P_S(u_{i-1},B,z)\right]
\end{equation}
and
\begin{eqnarray}
E_i&=&\int dz \int dB B\left[P_S(u_i,B,z)-P_S(u_{i-1},B,z)\right]\\ \nonumber
F_i&=&\int dz\int dB\int_{u_{i-1}}^{u_i}dA P_S(A,B,z)
\end{eqnarray}

\end{document}